\documentstyle[12pt,fbsty,epsf]{article}
\epsfverbosetrue
 
\font\tenrm=cmr10

 1
 1
 1

\def\ijmpa#1#2#3{  {\it Int. J. Mod. Phys. }{\bf A #1} (19#2) #3}

\def\npb#1#2#3{    {\it Nucl. Phys. }{\bf B\,#1} (19#2) #3}
\def\plb#1#2#3{    {\it Phys. Lett. }{\bf B\,#1} (19#2) #3}
\def\prd#1#2#3{    {\it Phys. Rev. }{\bf D\,#1} (19#2) #3}

\def\prl#1#2#3{    {\it Phys. Rev. Lett. }{\bf #1} (19#2) #3}
\def\ptp#1#2#3{    {\it Prog. Theor. Phys. }{\bf #1} (19#2) #3}
\def\rmp#1#2#3{    {\it Rev. Mod. Phys. }{\bf #1} (19#2) #3}
\def\zpc#1#2#3{    {\it Zeit. f\"ur Physik }{\bf C\,#1} (19#2) #3}

\def\ib#1#2#3{     {\it ibid. }{\bf #1} (19#2) #3}

\def\ltap{\ \raisebox{-.4ex}{\rlap{$\sim$}} \raisebox{.4ex}{$<$}\ }

\def\gsim{{~\raise.15em\hbox{$>$}\kern-.85em
          \lower.35em\hbox{$\sim$}~}}
\def\lsim{{~\raise.15em\hbox{$<$}\kern-.85em
          \lower.35em\hbox{$\sim$}~}}

\def\beq{\begin{equation}}
\def\eeq{\end{equation}}
\def\bea{\begin{eqnarray}}
\def\eea{\end{eqnarray}}

\newcommand{\nn}{\nonumber}
\newcommand{\f}{\frac}
\def\l{\lambda}
\newcommand{\llu}{\mbox{$V_{ub}V_{us}^*$}}
\newcommand{\llc}{\mbox{$V_{cb}V_{cs}^*$}}
\newcommand{\llt}{\mbox{$V_{tb}V_{ts}^*$}}
\newcommand{\hd}{{\overline D}}
\newcommand{\smallz}{{\scriptscriptstyle Z}} 
\newcommand{\smallw}{{\scriptscriptstyle W}} 
\newcommand{\smallh}{{\scriptscriptstyle H}} 
 
\newcommand{\mz}{M_\smallz}
\newcommand{\m}{\mu}
\newcommand{\mb}{m_b}
\newcommand{\mw}{M_\smallw}
\newcommand{\muw}{\mu_\smallw}
\newcommand{\mut}{\mu_t}
\newcommand{\mub}{\mu_b}
\newcommand{\mh}{m_\smallh}
\newcommand{\as}{\alpha_{\scriptscriptstyle S}}
\newcommand{\dnp}{\delta^{{\scriptscriptstyle NP}}}
\newcommand{\smallSL}{{\scriptscriptstyle SL}} 
\newcommand{\smallsm}{{\scriptscriptstyle SM}} 
\newcommand{\smallyy}{{\scriptscriptstyle YY}} 
\newcommand{\smallxy}{{\scriptscriptstyle XY}} 
\newcommand{\smallcp}{{\scriptscriptstyle CP}} 
\newcommand{\fudge}[1]{\noalign{\hbox{\parbox{\textwidth}{#1}}}\nonumber}
\def\tanb{\mbox{$\tan \! \beta\,$}}
\def\ctanb{\mbox{$\cot \! \beta\,$}}

\newcommand{\bsg}{\mbox{$b \to s \gamma\,$}}

\newcommand{\bbxsg}{\mbox{$\overline{B} \to X_s \gamma\,$}}

\newcommand{\bbrx}{\mbox{${\rm BR}(\overline{B}\to X_s\gamma)\,$}}

\newcommand{\barbbrx}{\mbox{${\rm BR}(B\to\overline{X_s}\gamma)\,$}}
\newcommand{\imx}{\rm{Im\,X}}
\newcommand{\rex}{\rm{Re\,X}}
\newcommand{\xx}{\mbox{$\rm{X}$}}
\newcommand{\yy}{\mbox{$\rm{Y}$}}
\newcommand{\zz}{\mbox{$\rm{Z}$}}
\newcommand{\xys}{{\rm{X\,Y}^\ast}}
\newcommand{\rxys}{{\rm Re}\,(\xx\,\yy^\ast)}
\newcommand{\ixys}{{\rm Im}\,(\xx\,\yy^\ast)}
\begin{document}

\begin{titlepage}
\noindent
\phantom{a}     \hfill         ZU--TH 31/97          \\
\phantom{a}     \hfill         BUTP--97/33          \\[7ex]

\begin{center}
{\Large
{\bf 2HDMs predictions for $\bbxsg$ in NLO QCD}}~\footnote{Work 
supported in part by Schweizerischer
 Nationalfonds}                                  \\[6ex]
{{\bf Francesca M.\ Borzumati$^{\,a}$ and
      Christoph Greub$^{\,b}$ }}                 \\[2ex]
$^a$ 
{\it Institut f\"ur Theoretische Physik, 
 Universit\"at Z\"urich                          \\
 Winterthurerstrasse 190, 
 8057 Z\"urich, Switzerland}                     \\[1.01ex]
$^b$ 
{\it Institut f\"ur Theoretische Physik,
 Universit\"at Bern                              \\
 Sidlerstrasse 5,  3012 Bern, Switzerland} 
                                                 \\[12ex]
\end{center}
{\begin{center} ABSTRACT \end{center}}
\vspace*{1mm}

\parbox{14.4cm}
{
\noindent
The decay $\bbxsg$ is studied at the Next to Leading Order in QCD in a
class of models containing at least two Higgs doublets and with only
one charged Higgs boson non--decoupled at low--energy. The two--loop
matching condition is calculated and it is found to agree with
existing results. The complete dependence of the Wilson coefficients
on the matching scale is given.  The size of the Next to Leading Order
corrections is extensively discussed. Results for branching ratios,
possible CP asymmetries and lower bounds on the charged Higgs mass are
presented when the convergence of the perturbative series appears fast
enough to yield reliable predictions. Regions in the parameter space
of these models where the Next to Leading Order calculation is still
not a good approximation of the final result for these observables are
singled out.
}
\vfill
\end{titlepage}

\thispagestyle{empty}

\setcounter{page}{1}

%%%%%%%%%%%%%%%%%%%%%%%%%%%%%%%%%%%%%%%%
%
\section{Introduction}
\label{intro}

\noindent 
It is well known that charged Higgs contributions to the branching
ratio for the decay \bbxsg, $\bbrx$, decouple very slowly from the
Standard Model (SM) one.  Hence, in the absence of any experimental
evidence for a charged Higgs boson, this decay may provide a powerful
tool to limit the range of unknown parameters in models where such a
particle is present and other --non standard-- contributions to
$\bbrx$ are subleading~\cite{Bbook,H-BBP}.

Supersymmetric models constitute, perhaps, the best motivated
extension of the SM where a second Higgs doublet, and therefore a
charged Higgs boson $H^\pm$, is necessary for internal
consistency~\cite{SUSYHiggs}. Simpler extensions where only Higgs
doublets are added to the SM are, however, important on their own
right. They are also excellent pedagogical tools to understand the
subtleties that the Next to Leading Order (NLO) calculation of $\bbrx$
entails, and which may be hidden in the SM results.

The simplest class of such extensions, with two Higgs doublets, is
usually denoted as 2HDMs. This class contains the well--known Type~I
and Type~II models in which the same or the two different Higgs fields
couple to up-- and down--type quarks.  For what concerns us here,
Multi--Higgs Doublet models can be included in this class, provided
only one charged Higgs boson remains light enough to be relevant for
the process $\bbxsg$.  This generalization allows a simultaneous study
of $\bbxsg$ in different models, including Type~I and Type~II, by a
continuous variation of the (generally complex) charged Higgs
couplings to fermions.  (No tree--level flavour violating neutral
couplings are assumed in the present paper.) It allows also a more
complete investigation of the question whether the measurement of
$\bbrx$ closes the possibility of a relatively light $H^\pm$ not
embedded in a supersymmetric model~\cite{IO}.

At present, a measurement of this decay rate 
by the CLEO Collaboration is available~\cite{CLEO}: 
\beq
 \bbrx = (2.32\pm 0.57\pm 0.35)\times 10^{-4}\,.
\label{CLEOres}
\eeq
There exists also a preliminary result by the 
ALEPH Collaboration with a larger central value~\cite{ALEPH}:
\beq
 \bbrx =(3.29\pm 0.71\pm0.68)\times 10^{-4} \,.
\label{ALEPHres}
\eeq
Adding statistical and systematical errors in 
quadrature, one obtains $90\%$ C.L. lower and upper
limits on $\bbrx$ of $1.22\times 10^{-4}$ and $3.42\times 10^{-4}$ from
the CLEO measurement and $1.68 \times 10^{-4}$, $4.90 \times 10^{-4}$ 
from the ALEPH result. The band of allowed values, 
corresponding to a more conservative estimate of the systematic error, 
is reported by CLEO to be 
$(1.0-4.2)\times 10^{-4}$~\cite{CLEO}.

The theoretical situation within the SM, is at the moment far better
settled than the experimental one. After the original observation that
QCD corrections to the decay $\bbxsg$ are a substantial fraction of
its rate~\cite{BBM}, a collective theoretical effort of almost a
decade has led to the determination of $\bbrx$ at the NLO in QCD, and
to a considerable reduction of the theoretical error.  For the Leading
Order (LO) calculation, several groups contributed to the evaluation
of the elements of the the anomalous dimension
matrix~\cite{GSW,CIUCHINI2} and provided phenomenological analyses
with partial inclusion of some NLO contributions (e.g. bremsstrahlung
corrections)~\cite{AG,LOWSCDEP,CIUCHINI1,NORESULTS}.  The two--loop
matching condition, needed for a complete NLO calculation, was first
obtained in~\cite{AYAO} and later confirmed
in~\cite{cGtH,INFRARED,CDGG}, using different techniques. The
two--loop corrections to the matrix elements were
calculated~\cite{GHW} and the determination of the ${\cal O}(\as^2)$
elements of the anomalous dimension matrix, started already since some
time~\cite{BURAS,MISIAK95}, has
been completed only recently~\cite{CMM,CMMlong}.  In addition,
non--perturbative contributions to $\bbrx$, scaling like
$1/m_b^2$~\cite{FALK,BIGI,LUKE} and $1/m_c^2$~\cite{VOLOSHIN}, were
also computed. The issue of the dependence of the branching 
ratio on scales, 
first raised at the LO level in~\cite{LOWSCDEP,NORESULTS}, was 
addressed for the NLO calculation, and discussed in detail, 
in~\cite{BKP}. In the SM, uncertainties due to sensitivity of 
$\bbrx$ on such scales are, 
undisputedly, rather 
small and the theoretical estimate for this observable 
suffers mainly from the large 
experimental errors of the input parameters.

It would be desirable to have similarly precise calculations 
also in extensions of the SM. Thus, more accurate experimental 
measurements, when available, could provide stringent
constraints on the free parameters of these models. 
We present here a detailed study of $\bbxsg$ at the NLO in QCD, in the
class of models specified above, aiming, in particular, to an
assessment of the reliability of the theoretical calculation.
Our results are, in general, less optimistic
than one could have foreseen. Indeed, we find unexpectedly large 
NLO corrections and scale dependences in the Higgs contributions to 
$\bbrx$, irrespectively of the value of the charged Higgs
couplings to fermions. This feature remains undetected in 
Type~II models, where the SM contribution to $\bbrx$ is always 
larger than, and in phase with, the Higgs contributions. 
It can, however, produce unacceptably large scale uncertainties for
certain ranges of these couplings and, at times, completely
ill--defined results. We single out combinations of couplings and 
of values of the charged Higgs mass, $\mh$, for which the branching
ratio can be reliably predicted. In this case, a comparison 
between theoretical and experimental results for $\bbrx$ 
allows to conclude that values of $\mh = {\cal O}(\mw)$ 
can be excluded, in general, only in Type~II models, but are 
otherwise allowed. 

Previous LO analyses had dealt with Type~I and Type~II 
models~\cite{GSW,HW,BHP,BBMR} and the generalized 
class of models considered here~\cite{KP,GNR}. LO calculations
are known to have large scale uncertainties. Consequently, they 
are not particularly good arenas to distinguish the 
quantitative differences between 
QCD corrections to charged Higgs and SM contributions. 
These differences stand out clearly at the NLO level. 
Two NLO calculations have been performed 
recently~\cite{CDGG,CRS}. They deal with Type~I and Type~II 
models and the issue of
scale uncertainty is addressed in~\cite{CDGG} for 
Type~II models. The fact that the NLO corrections 
to the Higgs contributions are large is, therefore, 
understandably missed, since the branching ratio is 
dominated by the SM contribution.
From the technical point of view, our calculation 
of the two--loop matching condition agrees 
with that reported in~\cite{CDGG,CRS}. 
We disagree, however, with the analytic dependence of 
$\bbrx$ on the the matching scale given in the 
literature for the SM~\cite{BKP} and 
for 2HDMs~\cite{CDGG}.

The reminder of this paper is organized as follows. In
Sect.~\ref{2hdms} we define the class of 2HDMs studied. In
Sect.~\ref{calculation} we outline the main steps of the calculation;
to keep this section readable, we relegate some parts of our results 
to appendices. In Sect.~\ref{results} we present our phenomenological
analysis of the decay $\bbxsg$ in the class of models
considered. After giving our branching ratio prediction for the SM, we
discuss, in~\ref{amplitudexy}, the size of the NLO corrections to the
charged Higgs contributions at the amplitude level. In
Sec.~\ref{realcouplings}, we give results for $\bbrx$, and discuss
their reliability, for various ranges of real couplings of the charged
Higgs to fermions. NLO branching ratios for Type~I and Type~II models
are in particular studied in Sec.~\ref{typeoneandtwo}, where lower
bounds for $\mh$, are given within Type~II
models.  In Sec.~\ref{complexcouplings}, we consider the decay
$\bbxsg$ when complex couplings are involved and we 
investigate the impact of their phases on
CP rate asymmetries of $\bbxsg$.  We give also specific examples of
couplings for which the theoretical prediction of the branching ratio
is reliable and compatible with the CLEO measurement, even for 
$\mh= {\cal O}(\mw)$.  Finally, our concluding remarks are in 
Sect.~\ref{conclusions}.

%%%%%%%%%%%%%%%%%%%%%%%%%%%%%%%%%%%%%%%%
%
\section{Two Higgs Doublet Models}
\label{2hdms}

\noindent 
Models with more than one Higgs doublet have generically a Yukawa 
Lagrangian of the form:
\beq 
-{\cal L} =
  h^d_{ij} \,{\overline{q'}_L}_i \,\phi_1 \, {d'_R}_j 
 +h^u_{ij} \,{\overline{q'}_L}_i \,{\widetilde \phi}_2 \,{u'_R}_j 
 +h^l_{ij} \,{\overline{l'}_L}_i \,\phi_3 \, {e'_R}_j 
 +{\rm h.c.}\,,
\label{yukpot}
\eeq
where $q'_L$, $l'_L$, $\phi_i$, ($i=1,2,3$) are SU(2) 
doublets (${\widetilde \phi_i} = i \sigma^2 \phi_i^*$); 
$u'_R$, $d'_R$, and $e'_R$ are SU(2) singlets
and 
$h^d$, $h^u$, and $h^l$ denote $3\times3$ Yukawa matrices. 
Gauge invariance
imposes the value of hypercharge $Y(\phi_i)=1/2$. 
Suitable discrete symmetries are usually invoked to forbid additional
terms like 
 $ h^d_{ij} \, {\overline{q'_L}}_i \, 
   \phi_2 \, {d'_R}_j $
which would induce flavour changing neutral couplings at the 
tree--level. Indeed, to avoid them altogether, it is sufficient to 
impose that no more than one Higgs doublet couples to the same 
right--handed field~\cite{GW}. 

When only two Higgs doublets are present, $\phi_1$ and $\phi_2$, 
it is in general $\phi_3 = \phi_1$ (or 
$\phi_3 = \phi_2$). Nevertheless, for the sake of generality, 
we leave the symbol $\phi_3$ distinct from the other two.
We indeed include in our discussion, as ``effective'' 
2HDMs, also models where an $n$--number of
sequential Higgs doublets is present and we assume 
that the additional charged Higg bosons other than the lightest 
one become heavy enough to decouple from our problem.

After rotating the fermion fields from the current eigenstate to the 
mass eigenstate basis, the charged Higgs component in eq.~(\ref{yukpot}) 
becomes:
\beq 
-{\cal L} =
 \f{{m_d}_i}{<\phi_1^0>}\,
  {\overline{u}_L}_j V_{ji} \,{d_R}_i  \,\phi_1^{\,+} 
-\f{{m_u}_i}{<\phi_2^0>}\,
  {\overline{u}_R}_i V_{ij} \,{d_L}_j  \,\phi_2^{\,+}
+\f{{m_l}_i}{<\phi_3^0>}\,{\overline{\nu}_L}_i\,{e_R}_i \,\phi_3^{\,+} 
+{\rm h.c.}\,,
\label{chargpot}
\eeq
where $V_{ji}$ are elements of the 
Cabibbo--Kobayashi--Maskawa (CKM) matrix. 
A further rotation of the charged Higgs fields to their physical 
basis through a unitary matrix $U$, 
\beq
\left(\begin{array}{l} 
 \phi_1^{\,+} \\ \phi_2^{\,+} \\ \phi_3^{\,+} \\ ... \\ \phi_n^{\,+}    
      \end{array} 
\right)   =  \  U  \,
\left(\begin{array}{l}  
 \phi^{\,+}   \\  H^+  \\ {\cal H}_1^{\,+}  \\  ...  \\
 {\cal H}_{n-2}^{\,+} 
      \end{array} 
\right)
\eeq 
yields the following Yukawa interaction for $H^+$:
\beq 
{\cal L} =\frac{\,g}{\sqrt{2}} \left\{
\left(\frac{{m_d}_i}{\mw}\right)
      \xx \,{\overline{u}_L}_j V_{ji}  \, {d_R}_i+
\left(\frac{{m_u}_i}{\mw}\right)
      \yy \,{\overline{u}_R}_i V_{ij}  \, {d_L}_j+
\left(\frac{{m_l}_i}{\mw}\right)
      \zz \,{\overline{\nu}_L}_i \, {e_R}_i 
                               \right\} H^+
 +{\rm h.c.}\,.
\label{higgslag}
\eeq
All fields ${\cal H}_i^{\,+}$ are supposed to be 
heavy enough to become irrelevant for phenomenology at the 
electroweak scale. In~(\ref{higgslag}), 
the symbols $\xx$ and $\yy$
are defined in terms of elements of 
the matrix $U$ ~\cite{WEIN,AST,BBG,KP,GNR}:
\beq
 \xx = -\f{U_{12}}{U_{11}}\,;   \hspace*{1cm}
 \yy = \f{U_{22}}{U_{21}}\,. 
\eeq
The symbol $\zz$ has a similar definition, 
i.e. $\zz =-{U_{32}}/{U_{31}}$ if $\phi_3 \ne \phi_1,\phi_2$ or 
coincides with $\xx$ ($-\yy$) if $\phi_3 = \phi_1$ 
($\phi_3 = \phi_2$). Notice that $\xx$, $\yy$, and $\zz$ are 
in general complex numbers and therefore potential sources of
CP violating effects~\cite{WEI,LAV}. Their values are only 
very loosely constrained by the requirement of perturbativity 
and low--energy processes such as the $B$--$\overline{B}$ 
mixing~\cite{GNR}. 
 
When only two doublets are considered, the diagonalization 
matrix $U$ is a $2\times 2$ orthogonal matrix 
\beq
\left(\begin{array}{rr} 
  \cos \beta & -\sin \beta  \\
  \sin \beta & \cos \beta  
      \end{array}
\right)  \,.  
\eeq
Although both doublets are present in the theory, one still has the 
freedom of selecting $\phi_1 = \phi_2$ in eq.~(\ref{yukpot}). This 
choice gives rise to the 2HDM of Type I, to be distinguished 
from the Type II in which both doublets contribute to the 
Yukawa interactions~\cite{HHS}. 
It is easy to see that in these two cases the couplings
$\xx$ and $\yy$ are real and given by:
\beq
\begin{array}{lll}
 X= -\ctanb, &\quad Y=  \ctanb         &\qquad {\rm (Type \ I)} \nn \\
 X= \phantom{-}\tanb,  &\quad Y=  \ctanb &\qquad {\rm (Type \ II)}\,.
\end{array}  
\label{typeItypeII}
\eeq 

Note that the coupling of the Goldstone boson $\phi^+$ 
to matter fields is independent of the number $n$ of 
Higgs doublets considered, and always equal to:
\beq
{\cal L}= -
\frac{\,g}{\sqrt{2}} \left\{
\left(\frac{{m_d}_i}{\mw}\right) \,
     {\overline{u}_L}_j V_{ji}   \, {d_R}_i-
\left(\frac{{m_u}_i}{\mw}\right) \,
     {\overline{u}_R}_i V_{ij}   \, {d_L}_j+
\left(\frac{{m_l}_i}{\mw}\right) \,{\overline{\nu}_L}_i \, {e_R}_i 
                               \right\} \phi^+
 +{\rm h.c.} \,,
\eeq
The calculation of the $\bbxsg$ decay rate in these models,
therefore, is modified simply by the addition of 
charged Higgs contributions to the usual SM one. The results
can be described in terms of $\mh$ and the two complex 
parameters $\xx$ and $\yy$. The presence of phases 
in the couplings $\xx$ and $\yy$ 
allows for CP asymmetries in the decay 
rate for $\bbxsg$.

%%%%%%%%%%%%%%%%%%%%%%%%%%%%%%%%%%%%%%%%%%%%%%%%%%%%%%%%%%%%
%
\section{The Calculation}
\label{calculation}

\noindent 
We use the framework of an effective low--energy theory with five
quarks, obtained by integrating out the heavy degrees of freedom,
which in the present case are the $t$--quark, the $W$--boson and 
the charged Higgs boson. As in the SM calculations, we only take
into account operators up to dimension six and we put $m_s=0$.
In  this approximation the   
effective Hamiltonian relevant for radiative $B$--decays
(with $ |\Delta B| = |\Delta S| =1$)  
\beq
 {\cal H}_{eff} = - \frac{4 G_F}{\sqrt{2}} \,V_{ts}^\star V_{tb} 
   \sum_{i=1}^8 C_i(\mu) {\cal O}_i(\mu)  
\eeq
consists precisely of the same operators ${\cal O}_i(\mu)$
used in the SM case, weighted by the Wilson coefficients 
$C_i(\mu)$. They read:
\beq
\begin{array}{llll}
{\cal O}_1 \,= &\!
 (\bar{s}_L \gamma_\mu T^a c_L)\, 
 (\bar{c}_L \gamma^\mu T_a b_L)\,, 
               &  \quad 
{\cal O}_2 \,= &\!
 (\bar{s}_L \gamma_\mu c_L)\, 
 (\bar{c}_L \gamma^\mu b_L)\,,   \\[1.002ex]
{\cal O}_3 \,= &\!
 (\bar{s}_L \gamma_\mu b_L) 
 \sum_q
 (\bar{q} \gamma^\mu q)\,, 
               &  \quad 
{\cal O}_4 \,= &\!
 (\bar{s}_L \gamma_\mu T^a b_L) 
 \sum_q
 (\bar{q} \gamma^\mu T_a q)\,,  \\[1.002ex]
{\cal O}_5 \,= &\!
 (\bar{s}_L \gamma_\mu \gamma_\nu \gamma_\rho b_L) 
 \sum_q
 (\bar{q} \gamma^\mu \gamma^\nu \gamma^\rho q)\,, 
               &  \quad 
{\cal O}_6 \,= &\!
 (\bar{s}_L \gamma_\mu \gamma_\nu \gamma_\rho T^a b_L) 
 \sum_q
 (\bar{q} \gamma^\mu \gamma^\nu \gamma^\rho T_a q)\,,  \\[1.002ex]
{\cal O}_7 \,= &\!
  \frac{e}{16\pi^2} \,{\overline m}_b(\mu) \,
 (\bar{s}_L \sigma^{\mu\nu} b_R) \, F_{\mu\nu}\,, 
               &  \quad 
{\cal O}_8 \,= &\!
  \frac{g_s}{16\pi^2} \,{\overline m}_b(\mu) \,
 (\bar{s}_L \sigma^{\mu\nu} T^a b_R)
     \, G^a_{\mu\nu}\,,
\end{array} 
\label{opbasis}
\eeq
where $T^a$ ($a=1,8$) are $SU(3)$ colour generators; 
$g_s$ and $e$, the strong and electromagnetic coupling constants. 
In eq.~(\ref{opbasis}), 
${\overline m}_b(\mu)$ is the running $b$--quark mass 
in the ${\overline{MS}}$ scheme at the renormalization scale $\mu$. 
Henceforth, ${\overline m}_q(\mu)$ and $m_q$ denote ${\overline{MS}}$ 
running and pole masses, respectively. To first order in 
$\as$, these masses are related through:
\beq
 {\overline m}_q(\mu) = {m_q} 
 \left(1 + \frac{\as(\mu)}{\pi} \, \ln \frac{m_q^2}{\mu^2} 
- \frac{4}{3} \frac{\as(\mu)}{\pi}\right) \, .
\label{polerunning}
\eeq
Note that the equations of motion have been used
when writing down the list of operators in 
eq.~(\ref{opbasis}). 
This is sufficient since we are interested in on--shell 
matrix elements~\cite{POLSIMMA} 
and since we choose to perform the matching 
by comparing  on--shell
amplitudes obtained in the effective theory with the corresponding
amplitudes in the full theory. 
The reader who is interested
in doing off--shell matching (and therefore working in a (larger)
off--shell operator basis) is referred to ref. \cite{CDGG}.

Working to NLO precision means that one is resumming all the
terms of the form $\as^n(\mb) \, \ln^n (\mb/M)$, as well as
$\as(\mb) \, \left(\as^n(\mb) \, \ln^n (\mb/M)\right)$,
where $M$ stands for one of the heavy masses $\mw$, $m_t$ or $m_H$.
This is achieved by performing the following 3 steps: %\newline

\noindent 
1) One matches the full standard model theory
with the effective theory at a scale $\muw$, of order $M$. At this scale,
the matrix elements of the operators  in the 
effective theory lead to the  same logarithms  as the 
calculation in the full theory. 
Consequently, the Wilson coefficients 
$C_i(\muw)$ only pick up ``small'' QCD corrections,
which can be calculated in fixed order perturbation theory.
In the NLO program, the matching has to be worked out at the 
$O(\as)$ level.  %\newline

\noindent 
2) The evolution of these Wilson coefficients from 
$\m=\muw$ down to $\m = \mub$, where $\mub$ is of the order of $m_b$,
is obtained by solving the appropriate Renormalization
Group Equation (RGE). As the matrix elements of the operators 
evaluated at the low scale
$\mub$ are free of large logarithms, the latter are contained in resummed
form in the Wilson coefficients. For a NLO calculation, this step
has to be performed using the anomalous dimension matrix up 
to order $\as^2$. %\newline

\noindent 
3) The corrections to the matrix elements of the operators
$\langle s \gamma | {\cal O}_i (\mu) |b \rangle$ at the scale  
$\mu = \mub$ have to be calculated to order $\as$ precision.

\noindent 
The charged Higgs boson enters the NLO calculation only via 
step 1). The Higgs boson contribution to the matching condition  
is obtained in the same way as the SM 
one~\cite{cGtH}; therefore, we do not 
repeat any technical details. A general remark, however,
is in order. In the procedure described above 
all heavy particles ($W$--boson, $t$--quark
and charged Higgs) are integrated out simultaneously at the scale
$\muw$. In the context of a 
NLO calculation this should be a reasonable 
approximation provided $\mh$ is of the same order of magnitude
as $\mw$ or $m_t$.     

Before giving the results of the three steps listed above,
we should briefly mention  that
instead of the original Wilson coefficients $C_i(\mu)$ is it 
convenient to use
certain linear combinations of them,
the so--called ``effective Wilson'' coefficients
$C_i^{\,{\rm eff}}(\mu)$ introduced in~\cite{NORESULTS,CMM}:
\begin{eqnarray}
\label{cieff}
C_{i}^{\,{\rm eff}}(\mu) &=& C_i(\mu) \,, \quad (i=1,...,6) \,, \nn \\
C_7^{\,{\rm eff}}(\mu) &=& C_7(\mu) +\sum_{i=1}^6 y_i C_i(\mu) \,, \quad
C_8^{\,{\rm eff}}(\mu) =   C_8(\mu) +\sum_{i=1}^6 z_i C_i(\mu) \,, \quad
\end{eqnarray} 
where $y_i$ and $z_i$ are defined in such a way
that the leading oder matrix elements 
$\langle s \gamma |{\cal O}_i|b \rangle$ and
$\langle s g      |{\cal O}_i|b \rangle$ ($i=1,6$) are 
absorbed in the leading orders terms
in $C_7^{\,{\rm eff}}(\mu)$ and $C_8^{\,{\rm eff}}(\mu)$. 
The explicit values of $\{y_i\}$ and $\{z_i\}$, 
$y=(0,0,-\frac{1}{3},-\frac{4}{9},-\frac{20}{3},-\frac{80}{9})$,
$z=(0,0,1,-\frac{1}{6},20,-\frac{10}{3})$  
were obtained in ref.~\cite{CMM} 
in the $\overline{MS}$ scheme with
fully anticommuting $\gamma_5$ (also used in the
present paper).

%%%%%%%%%%%%%%%%%%%%%%%%%%%%%%%%%%%%%%%%%%%%%%%%%
%
\subsection{NLO Wilson coefficients at the matching scale $\muw$:
            $C_i^{\,{\rm eff}}(\muw)$}
\label{wcmuw}

\noindent 
To give the results for the effective Wilson coefficients
$C_i^{\,{\rm eff}}$ 
at the matching scale $\muw$ in a compact form,
we write:
\beq
 C^{\,{\rm eff}}_i(\muw) =  C^{0,\,{\rm eff}}_i(\muw)
             + \f{\as(\muw)}{4\pi} C^{1,\,{\rm eff}}_i(\muw) \,.
\label{effcoeff}
\eeq
The LO Wilson coefficients at this scale are well 
known~\cite{INAMILIM,HW}. We decompose them in such a way to 
render explicit their dependence on the couplings 
$\xx$ and $\yy$:
\bea 
 C^{0,\,{\rm eff}}_2(\muw)  & = &  1                       \nn \\  
 C^{0,\,{\rm eff}}_i(\muw)  & = &  0 
  \hspace*{6.5truecm} (i=1,3,4,5,6)                        \nn \\
 C^{0,\,{\rm eff}}_7(\muw)  & = &   C_{7,\smallsm}^0
  +\vert \yy \vert^2  \, C_{7,\smallyy}^0
  +         (\xys)    \, C_{7,\smallxy}^0    \nn \\
 C^{0,\,{\rm eff}}_8(\muw)  & = &   C_{8,\smallsm}^0
  +\vert \yy \vert^2  \, C_{8,\smallyy}^0
  +         (\xys)    \, C_{8,\smallxy}^0 \,.   
\label{coeffLO}
\eea
The coefficients 
$C_{7,\smallsm}^0(\muw)$ and $C_{8,\smallsm}^0(\muw)$ are
functions of $x=m_t^2/\mw^2$, while 
$C_{7,j}^0(\muw)$ and
$C_{8,j}^0(\muw)$ ($j={\mbox{\small{YY,\,XY}}}$) 
are functions of $y=m_t^2/m_H^2$; their explicit forms are given in
Appendix~\ref{defineCoeff}. Note that there is no {\it explicit}
dependence of the matching scale $\muw$ in these functions. Whether
there is an {\it implicit} $\muw$--dependence via the $t$--quark mass
depends on the precise definition of this mass which has to be
specified when going beyond leading logarithms. If one chooses for
example to work with $\overline{m_t}(\muw)$, then there is such an
implicit $\muw$--dependence of the lowest order Wilson coefficient; in
contrast, when working with the pole mass $m_t$ there is no such
dependence. We choose to express our (NLO) results in 
terms of the pole mass $m_t$. 

The NLO pieces $C_i^{1,\,{\rm eff}}(\muw)$ 
of the Wilson coefficients have an explicit 
dependence on the matching scale $\muw$ 
and for $i=7,8$ they also explicitly depend on the actual
definition of the $t$--quark mass. 
Initially, when the heavy particles are integrated out, it is
convenient to work out the matching conditions
$C_i^{1,\,{\rm eff}}(\muw)$ for $i=7,8$ 
in terms of ${\overline m_t}(\muw)$.
Using eq.~(\ref{polerunning}), it is then   
straightforward to get  the corresponding 
result expressed in terms of the pole mass $m_t$.  
As in the LO case we give them 
in a form where the dependence of the
couplings $\xx$ and $\yy$ is explicit:
\bea 
 C_ 1^{1,\,{\rm eff}}(\muw) & = &
 15 + 6 \ln\frac{\muw^2}{\mw^2}                            \nn \\
 C_ 4^{1,\,{\rm eff}}(\muw) & = &
 E_0 + \frac{2}{3} \ln\frac{\muw^2}{\mw^2}  
  +\vert \yy\vert^2  \, E_H                                   \nn \\
 C_ i^{1,\,{\rm eff}}(\muw) & = &   0  
  \hspace*{6.5truecm}  (i=2,3,5,6)                          \nn \\[1.5ex]
 C_ 7^{1,\,{\rm eff}}(\muw) & = &
 C_{7,\smallsm}^{1,\,{\rm eff}}(\muw)
  +\vert \yy\vert^2 \, C_{7,\smallyy}^{1,\,{\rm eff}}(\muw)
  +        (\xys)   \, C_{7,\smallxy}^{1,\,{\rm eff}}(\muw)  \nn \\[1.5ex]
 C_ 8^{1,\,{\rm eff}}(\muw) & = &
 C_{8,\smallsm}^{1,\,{\rm eff}}(\muw)
  +\vert \yy\vert^2 \, C_{8,\smallyy}^{1,\,{\rm eff}}(\muw)
  +       (\xys)    \, C_{8,\smallxy}^{1,\,{\rm eff}}(\muw) \,,
\label{coeffNLO}
\eea
where for $i=7,8$ the three terms on the right hand side
can be written in the form
\bea
C_{i,\smallsm}^{1,\,{\rm eff}}(\muw)                          & = & 
  W_{i,\smallsm} + M_{i,\smallsm} \ln\frac{\muw^2}{\mw^2} 
                   + T_{i,\smallsm} 
  \left( \ln\frac{m_t^2}{\muw^2} -\frac{4}{3} \right)   \nn \\
C_{i,\smallyy}^{1,\,{\rm eff}}(\muw)                          & = &  
  W_{i,\smallyy} + M_{i,\smallyy} \ln\frac{\muw^2}{\mh^2}  
                 \,+ T_{i,\smallyy} 
  \left( \ln\frac{m_t^2}{\muw^2} - \frac{4}{3} \right)   \nn \\
C_{i,\smallxy}^{1,\,{\rm eff}}(\muw)                          & = &  
  W_{i,\smallxy} + M_{i,\smallxy} \ln\frac{\muw^2}{\mh^2} 
                 \,+ T_{i,\smallxy} 
   \left( \ln\frac{m_t^2}{\muw^2} - \frac{4}{3} \right) \,.
\label{coeffNLOa}
\eea
Note that in eq.~(\ref{coeffNLOa}) 
the $W_{i,j}$--  and the $M_{i,j}$ ($j={\mbox{\small{SM,\,XY,\,YY}}}$)
terms would be the full result when working in terms of
${\overline m}_t(\muw)$. The $T_{i,j}$ terms result when expressing
${\overline m}_t(\muw)$ in terms of the pole mass $m_t$ 
in the corresponding lowest
order coefficients. Thus, for $i=7,8$, 
the $T_{i,j}$--quantities are
\beq
\label{Tterm}
T_{i,SM} = 8 \, x \,
 \f{\partial C_{i,\smallsm}^{0,\,{\rm eff}}(\muw)}
                        {\partial x} \,, \quad
T_{i,j} = 8 \, y \, 
 \f{\partial C_{i,j}^{0,\,{\rm eff}}(\muw)}{\partial y} \,,
   \quad (j={\mbox{\small{XY,\,YY}}}) \,.
\eeq   
Notice that if one worked with the running $t$--quark mass 
${\overline m_t}(\mut)$
normalized at the scale $\mut$ instead of the pole mass $m_t$,
the third terms on the right hand sides of eqs.~(\ref{coeffNLOa}) 
would have
to be replaced by $T_{i,j} \ln ({\mut^2}/{\muw^2})$ 
$(j={\mbox{\small{SM,\,YY,\,XY}}})$. 
The functions 
$W_{i,j}$, $M_{i,j}$ and $T_{i,j}$ 
($j={\mbox{\small{SM,\,XY,\,YY}}}$), 
together with $E_0$ and $E_H$, 
are listed in Appendix~\ref{defineCoeff}. Our results for 
$C_7^{1,\,{\rm eff}}(\muw)$
and $C_8^{1,\,{\rm eff}}(\muw)$ 
agree with those in refs.~\cite{CRS} and~\cite{CDGG}, when 
taking into account that
the latter results are expressed in terms of
${\overline m_t}(\muw)$. We disagree, however, with the form 
of the coefficients $C_{1}^{1,\,{\rm eff}}(\muw)$    
and $C_{4}^{1,\,{\rm eff}}(\muw)$, whose matching scale
dependence is forgotten in~\cite{BKP,CDGG}. As it will be seen later 
on, this neglect has consequences in the determination of 
the matching scale which yields the lowest estimate of 
$\bbrx$.

%%%%%%%%%%%%%%%%%%%%%%%%%%%%%%%%%%%%%%%%%%%%
%
\subsection{NLO Wilson coefficients at the low--scale $\mub$:
            $C_i^{\,{\rm eff}}(\mub)$}
\label{wcmub}

\noindent 
The evolution from the matching scale $\muw$ down to the
low--energy scale $\mub$
is described by the RGE:
\beq
\label{RGE}
 \mu \frac{d}{d\mu} C_i^{{\rm eff}}(\mu) = 
 C_j^{\,{\rm eff}}(\mu) \, \gamma_{ji}^{\,{\rm eff}} (\mu) \,.
\eeq
The initial conditions $C_i^{\,{\rm eff}}(\muw)$ for this equation
are given in 
Section~\ref{wcmuw},
and the anomalous dimension matrix $\gamma_{ij}^{\,{\rm eff}} $ 
is given in Appendix~\ref{anomalousdm} up to order
$\as^2$, which is the precision 
needed for a NLO calculation. 
The solution of eq.~(\ref{RGE}), obtained through the
procedure described in~\cite{BBL}, yields for the 
coefficient
\beq
\label{wilsonsplit}
 C_i^{{\rm eff}}(\mub) = C_i^{0,\,{\rm eff}}(\mub) + 
 \frac{\as(\mub)}{4\pi} \, C_i^{1,\, {\rm eff}}(\mub) \,, 
\eeq
the LO term:
\beq 
 C^{0,\,{\rm eff}}_7(\mub)   =
  \eta^\f{16}{23}  C^{0,\,{\rm eff}}_7(\muw) 
 +\f{8}{3} \left(\eta^\f{14}{23} -\eta^\f{16}{23}\right)
                   C^{0,\,{\rm eff}}_8(\muw) 
 + \sum_{i=1}^8 h_i \,\eta^{a_i} \,C^{0,\,{\rm eff}}_2(\muw) \,,
\label{runc70}
\eeq
and the NLO one:
\bea     
 C^{1,\,{\rm eff}}_7(\mub) & = & 
  \eta^{\f{39}{23}} C^{1,\,{\rm eff}}_7(\muw) + 
 \f{8}{3} \left( \eta^{\f{37}{23}} - \eta^{\f{39}{23}} \right) 
      C^{1,\,{\rm eff}}_8(\muw)                         \nn \\[1.05ex] 
                     &   &
+\left( \f{297664}{14283}   \eta^{\f{16}{23}}
       -\f{7164416}{357075} \eta^{\f{14}{23}} 
       +\f{256868}{14283}   \eta^{\f{37}{23}}
       -\f{6698884}{357075} \eta^{\f{39}{23}}
 \right) C^{0,\,{\rm eff}}_8(\muw)                      \nn \\[1.1ex]
                     &   &
+\, \f{37208}{4761} \left(\eta^{\f{39}{23}} -\eta^{\f{16}{23}} \right) 
          C^{0,\,{\rm eff}}_7(\muw)                     \nn \\
                     &   &
 + \sum_{i=1}^8  \left(
   e_i \,\eta \,C^{1,\,{\rm eff}}_4(\muw)
 + (f_i + k_i \eta) \,C^{0,\,{\rm eff}}_2(\muw)
 +  l_i \,\eta\, C^{1,\,{\rm eff}}_1(\muw) \right) \eta^{a_i} \,,
\label{runc7eff1}
\eea
The symbol $\eta$ is defined as
$\eta=\as(\muw)/\as(\mub)$; 
the vectors $a_i$, $h_i$, $e_i$, $f_i$, $k_i$ and $l_i$
are listed in Appendix~\ref{runnnumbers}.
Notice that eq.~(\ref{runc7eff1}) can be used in this form for all
models in which the same set of coefficients $C_i(\muw)$ are non
vanishing. It is more general than the corresponding eq.~(21) given
in~\cite{CMM}, which can be used only when the matching
scale $\muw$ is fixed at the value $\mw$. The use made of that
equation in~\cite{BKP}, where the dependence of $\bbrx$ on the
matching scale is studied, is inappropriate.

As far as the other Wilson coefficients are concerned, they
are only needed to LO precision in the complete NLO analysis 
of $\bbrx$. In this precision, the Wilson coefficients
of the four--Fermi operators ($i=1,...,6$)
are the same as in the SM. As the coefficients 
$C_3^{\,{\rm eff}}(\mub),...,C_6^{\,{\rm eff}}(\mub)$ are
numerically much smaller than $C_2^{\,{\rm eff}}(\mub)$, we 
neglect contributions proportional to these
small Wilson coefficients in the amplitude for $\bbrx$ and
list here only the LO expressions for 
$C_1^{\,{\rm eff}}(\mub)$,
$C_2^{\,{\rm eff}}(\mub)$,
and $C_8^{\,{\rm eff}}(\mub)$:
\bea
\label{othercoeff}
 C^{0,\,{\rm eff}}_1(\mub) & = & \left( \eta^{\frac{6}{23}} -
\eta^{-\frac{12}{23}} \right) \, 
 C^{0,\,{\rm eff}}_2(\muw) \nn \\
 C^{0,\,{\rm eff}}_2(\mub) & = & \left( \frac{2}{3} \eta^{\frac{6}{23}} +
\frac{1}{3} \eta^{-\frac{12}{23}} \right) \, 
 C^{0,\,{\rm eff}}_2(\muw) \nn \\
 C^{0,\,{\rm eff}}_8(\mub) & = &
  \eta^\f{14}{23}  C^{0,\,{\rm eff}}_8(\muw) 
 + \sum_{i=1}^5 h_i^\prime \,\eta^{a^\prime_i}
  \,C^{0,\,{\rm eff}}_2(\muw) \,,
\eea
When discussing the contributions due to the charged Higgs boson, it is
convenient to split 
the Wilson coefficients at the scale $\mub$ 
into the contributions 
$C_{i,\smallsm}^{\,{\rm eff}}(\mub)$,
$C_{i,\smallyy}^{\,{\rm eff}}(\mub)$, and 
$C_{i,\smallxy}^{\,{\rm eff}}(\mub)$:
\beq
 C_{i}^{\,{\rm eff}}(\mub)  = 
  C_{i,\smallsm}^{\,{\rm eff}}(\mub) 
+ \vert \yy \vert^2 C_{i,\smallyy}^{\,{\rm eff}}(\mub) 
+ \left(\xys \right) C_{i,\smallxy}^{\,{\rm eff}}(\mub) \,.
\label{wcdecomp} 
\eeq
As the solution~(\ref{runc7eff1}) of the renormalization group 
equation~(\ref{RGE}) is linear in the initial
conditions, this splitting is given in an obvious way 
in terms of the corresponding splitting
at the matching scale $\muw$, presented in Section~\ref{wcmuw}.

When calculating NLO results in the  numerical analyses
in Section~\ref{results},
we use the NLO expression for the strong coupling constant:
\beq
\label{aqcd}
\as(\mu) = \frac{\as(\mz)}{v(\mu)} \, 
\left[ 1 - \frac{\beta_1}{\beta_0} \frac{\as(\mz)}{4\pi} \,
\frac{\ln v(\mu)}{v(\mu)} \right] \,,
\eeq 
with
\beq
v(\mu) = 1 - \beta_0 \frac{\as(\mz)}{2\pi} \, \ln \left(
\frac{\mz}{\mu} \right) \,,
\eeq
where $\beta_0=\frac{23}{3}$ and $\beta_1=\frac{116}{3}$.
However, for LO results 
we always use the LO expression for $\as(\mu)$, i.e.,
$\beta_1$ is put to zero in eq.~(\ref{aqcd}).

%%%%%%%%%%%%%%%%%%%%%%%%%%%%%%%%%%%%%%%%%%%%
%
\subsection{Branching ratio for $\bbrx$}
\label{brcal}

\noindent 
We first give the formulas for the quark decay $b \to X_s \gamma$
and discuss the meson decay $\bbxsg$ later.
In a NLO calculation, $b \to X_s \gamma$ involves the subprocesses
$b \to s \gamma$ (including virtual corrections) and 
$b \to s \gamma g$, i.e., the gluon bremsstrahlung process.
The amplitude for the first can be written as
\beq
\label{amplitude}
{\cal A}(b \to s \gamma) = - 
 \frac{4 G_F}{\sqrt{2}} \, V_{ts}^\star V_{tb} \, \hd \,
 \langle s \gamma |{\cal O}_7|b \rangle_{tree} \,,
\eeq
where the reduced amplitude $\hd$ is 
\beq
 \hd =  C^{0,\,{\rm eff}}_7(\mub)
  +\frac{\as(\mub)}{4 \pi} 
    \left(C_7^{1,\,{\rm eff}}(\mub) + V(\mub) \right)\,.
\label{drewrite}
\eeq
The symbol $V(\mub)$ is defined as: 
\beq
  V(\mub) = 
   \sum_{i=1}^8 C_i^{0,\,{\rm eff}}(\mub) 
   \left[r_i+\frac{1}{2}\gamma_{i7}^{0,\,{\rm eff}} 
             \ln \frac{m^2_b}{\mu^2_b} 
   \right] 
 - \f{16}{3} C_7^{0,\,{\rm eff}}(\mub) \,.
\label{vdefine}
\eeq
In writing eq.~(\ref{amplitude}) we directly converted the running mass 
factor $\overline{m_b}(\mu_b)$, which appears in the definition
of the operator ${\cal O}_7$ in eq.~(\ref{opbasis}), into the pole
mass $m_b$ by making use of eq.~(\ref{polerunning}). 
This conversion is absorbed into the function $V(\mub)$ and consequently
the symbol  
$ \langle s \gamma |{\cal O}_7|b \rangle_{tree}$ is the tree-level
matrix element of the operator
${\cal O}_7$, where the running mass factor $\overline{m_b}(\mu_b)$ is
understood to be replaced by the pole mass $m_b$.   
In the previous
literature, this procedure was done in two step. First, 
$\overline{m_b}(\mu_b)$ was expressed in terms of $\overline{m_b}(m_b)$
and then in turn $\overline{m_b}(m_b)$ was converted into the pole mass
$m_b$. This last step brought into the game the quantity $F$ (see 
e.g. ref.~\cite{GHW}). 
The elements $\gamma_{i7}^{0,\,{\rm eff}}$ 
of the anomalous dimension matrix 
and the virtual correction functions 
$r_i$ in~(\ref{vdefine}) are given in 
Appendix~\ref{anomalousdm} and~\ref{specifyD}, respectively.
Note, that some parts of the bremsstrahlung contributions
associated with ${\cal O}_7$ are effectively transferred 
to $r_7$ as detailed in ref.~\cite{GHW}.

A splitting analogous to that in~(\ref{wcdecomp}) holds also 
for the reduced amplitude $\hd$:
\beq
 \hd = \hd_{\smallsm} + \vert \yy \vert^2 \hd_{\smallyy} 
                  + \left(\xys \right) \hd_{\smallxy}\,.
\label{decompd} 
\eeq  
as well as for the function $V(\mub)$. 

From ${\cal A}(b \to s \gamma)$ in eq.~(\ref{amplitude}) the 
decay width $\Gamma(b \to s \gamma)$ is easily obtained to be
\beq
\Gamma (b \to s \gamma) = \frac{G_F^2}{32\pi^4} \, 
 \vert V_{ts}^\star V_{tb}\vert ^2 \alpha_{em} \,
 m_b^5 \, |\hd|^2 \,.
\eeq
In the numerics we discard those
terms in $|\hd|^2$ which are explicitly of order $\as^2$.
Note, however, that there are implicit higher order terms which
are retained. Such terms arise for example because we evaluate
the quantity $\eta=({\as(\muw)}/{\as(\mub)})$ using the NLO
expression for $\as$ also in those Wilson coefficients 
which are needed only in LO precision. 

To obtain the inclusive rate for $b \to X_s \gamma$ consistently at the
NLO level, we have to take into account the
bremsstrahlung contributions \cite{AG,POTT}. 
The corresponding decay width
$\Gamma(b \to s \gamma g)$ is of the form
\beq
\label{brems}
\Gamma(b \to s \gamma g)= \frac{G_F^2}{32 \pi^4} \, 
 \vert V_{ts}^\star V_{tb}\vert ^2 
 \alpha_{em} \, m_b^5 \, A \,,
\eeq
where $A$ is
\beq
\label{aterm}
A = \frac{\as(\mub)}{\pi} \, \sum_{i,j=1;i \le j}^8 \, 
  {\rm Re} \left\{ C_i^{0,\,{\rm eff}}(\mub) \,
            \left[ C_j^{0,\,{\rm eff}}(\mub)\right]^*
 \, f_{ij} \right\} \,.
\eeq
As in the virtual contributions we put $C_{i}^{0,\,{\rm eff}}=0$ for 
$i=3,...,6$. In contrast to ref.~\cite{CMM}, we do not introduce
a cut--off when the photon gets soft. In order to cancel the infrared
singularity, which appears in this case, we include 
as in ref.~\cite{GHW} the
virtual photonic correction to the process $b \to s g$, which we
absorb into the quantity $f_{88}$ (see eq. (B.9) in ref.~\cite{GHW}). 
Note also, that in our approach 
the term $f_{77}$ is already absorbed into the function $r_7$. 
The non--vanishing $f_{ij}$ terms, 
which have to be taken into account
explicitly in eq.~(\ref{aterm}), are listed in 
Appendix~\ref{defineeffij}.

In order to get the decay width for the meson decay $\bbxsg$ 
we take into account the non--perturbative corrections 
which scale like $1/m_b^2$~\cite{FALK} and those which scale like 
$1/m_c^2$~\cite{VOLOSHIN}.  The decay width $\Gamma(\bbxsg)$ then 
reads:
\bea
\label{widthrad}
\lefteqn{
\Gamma(\bbxsg) \ = \ \frac{G_F^2}{32\pi^4}  
 \vert V_{ts}^\star V_{tb}\vert ^2 
 \alpha_{em} \, m_b^5  \hspace*{4.2cm} }    \nn \\[1.2ex]
& & 
\left\{ 
 |\hd|^2 + A + 
 \frac{\dnp_\gamma}{m_b^2} 
 \vert C_7^{0,\,{\rm eff}}(\mub) \vert^2 + 
  \frac{\dnp_c}{m_c^2}  
 {\rm Re} 
 \left\{
 \left[C_7^{0,\,{\rm eff}}(\mub)\right]^* \! 
 \left(C_2^{0,\,{\rm eff}}(\mub)
    \!-\!\frac{1}{6} C_1^{0,\,{\rm eff}}(\mub) \!\right)\! 
 \right\}\!
\right\}.
\quad \quad
\eea 
Since we work in the new operator basis, 
 a contribution to the term in $1/m_c^2$ comes not only from 
 ${\cal O}_2$, as it was incorrectly assumed in~\cite{CDGG},
 but also from ${\cal O}_1$. 
The non--perturbative quantities 
$\dnp_\gamma$~\cite{FALK} and $\dnp_c$~\cite{VOLOSHIN} 
in~(\ref{widthrad}) are: 
\beq
\dnp_\gamma = 
\frac{\lambda_1}{2} - \frac{9 \lambda_2}{2} \,, \quad \quad
\dnp_c = - \frac{ \lambda_2}{9} \,,
\eeq
where $\lambda_1$ and $\lambda_2$ parametrize the kinetic energy of
the $b$--quark and the  
chromomagnetic interactions, respectively.
Their values are: $\lambda_1 = -0.5\,$GeV$^2$ and 
$\lambda_2 = -0.12\,$GeV$^2$. Of these two parameters,
the first has larger uncertainties that the second one. 
As it will appear from the final formula for 
$\bbrx $, the overall $\l_1 $ dependence cancels 
to a large extent.

The branching ratio is then obtained as
\beq
\label{br}
\bbrx =  \frac{\Gamma(\bbxsg)}{\Gamma_{\smallSL}} 
 \, {\rm BR}_{\smallSL}\,,
\eeq
where ${\rm BR}_{\smallSL}$ is the measured semileptonic branching
ratio and the  
semileptonic decay width $\Gamma_{\smallSL}$ is given by:
\beq
\label{semileptonic}
\Gamma_{\smallSL} =
 \frac{G_F^2}{192\pi^3} |V_{cb}|^2 \,  m_b^5 \, g(z) \, 
 \left( 1 - \frac{2 \as(\overline{\mu}_b)}{3\pi} \, f(z) + 
 \frac{\dnp_{\smallSL}}{m_b^2} \right) \,; \quad
z=\frac{m_c^2}{m_b^2} \,.
\eeq 
The phase space function $g(z)$ and the (approximated) 
QCD--radiation function $f(z)$ \cite{CABIB} in~(\ref{semileptonic})
are:
\beq
\label{gf}
g(z) =1-8z+8z^3-z^4-12z^2 \ln z \,, \quad \quad 
f(z) = \left(\pi^2 -\f{31}{4} \right) (1-\sqrt{z})^2 +\f{3}{2} \,,
\eeq
and the non--perturbative correction $\dnp_{\smallSL}$ 
reads~\cite{BIGI,LUKE}:
\beq
 \dnp_{\smallSL} = \frac{\lambda_1}{2} + \frac{3\lambda_2}{2} \,
 \left[ 1 -4 \frac{(1-z)^4}{g(z)} \right]   \,.
\eeq 
In eq.~(\ref{semileptonic}) $\overline{\mu}_b$ is the renormalization
scale relevant to the semileptonic process, which is a priori
different from the renormalization scale $\mub$ 
in the radiative decay, as
stressed in~\cite{BKP}. However, as pointed out in
ref.~\cite{GHHawaii}, the identification 
${\overline \mu}_b=\mu_b$
turns out to be more conservative. We therefore use
${\overline \mu}_b=\mu_b$ in the numerical analysis.
Furthermore, in the evaluation of eq.~(\ref{br}),
we do not expand $1/\Gamma_{\smallSL}$ in powers of $\as$.

%%%%%%%%%%%%%%%%%%%%%%%%%%%%%%%%%%%%%%%%%%%%
%
\section{Results}
\label{results}

\noindent 
We discuss in this section
results and theoretical uncertainties of the NLO
calculation of $\bbrx$ in 2HDMs. 
As it will be shown, these uncertainties
can be very large, in contrast with what is found in the
SM. For reference, we give our SM result:
\beq
 \bbrx = \left(
       3.57 \ \pm^{\,0.01}_{\,0.12} \ {\rm (\mub)}
            \ \pm^{\,0.00}_{\,0.08} \ {\rm (\muw)}
            \ \pm^{\,0.29}_{\,0.27} \ {\rm (param)}
         \right) \times 10^{-4}  \,. 
\label{resultbrsm}
\eeq 
The central value $3.57 \times 10^{-4}$ is obtained for 
$\mub=4.8\,$GeV, $\muw = \mw$ and the central values of the input 
parameters listed in Table~\ref{inputpar} of 
Appendix~\ref{InputsScales}.
The low--scale variation in the interval
$[2.4,9.6]\,$GeV gives the maximum value for $\bbrx$ at 
$\mub=4.2\,$GeV and the minimum at $\mub=9.6\,$GeV. The matching 
scale dependence of the branching ratio is monotonically decreasing 
for increasing $\muw$:
from the central value in~(\ref{resultbrsm}),  
$\bbrx$ is reduced by $2\%$ at $\muw=m_t$. 
The value $3.57 \times 10^{-4}$ in eq.~(\ref{resultbrsm}) 
reduces to $3.46 \times 10^{-4} $ when the 
factor $1/\Gamma_{\smallSL}$ in eq.~(\ref{br}) is expanded 
in $\as$. 

Our results for the branching ratio $\bbrx$ in the class of 2HDMs 
considered, is parametrized in terms of $\{\xx,\yy,\mh\}$. 
We limit the range of this three--dimensional parameter space 
as follows. We fix $\yy$ to small real values of 
${\cal O}(1)$ and scan $\xx$ in the complex plane without ever
violating the perturbativity requirement
on $\xx$ and $\yy$ discussed in~\cite{GNR}. 
As explained in~\cite{BD}, the value of $\mh$ in 2HDMs can be as 
low as the LEP~I lower bound of $45\,$GeV. 
We do not strictly apply cuts on $\{\xx,\yy,\mh\}$ due to the 
measurement of $R_b$ or of other processes like 
$B$--${\overline B}$ mixing with virtual exchange of charged Higgs 
bosons, which 
could remove some corners of the parameter space considered. Our 
interest is more to show some theoretical 
features of the NLO calculation of $\bbrx$ 
in regions as wide as possible and to discuss constraints due to 
the decay $\bbxsg$ itself. 
For the numerical evaluations, unless otherwise specified, we
use the central value of the input parameters listed in 
Table~\ref{inputpar} in Appendix~\ref{InputsScales}. The values of 
the matching and low--scale, chosen respectively
in the intervals $[\mw, {\rm Max}(m_t,\mh)]$ and $[2.4,9.6]\,$GeV,
are explicitly given for each result presented. 

In order to investigate the reliability of the NLO prediction for 
the branching ratio, we study the term $ \vert \hd \vert ^2$, 
which dominates the expression within curly brackets 
in~(\ref{widthrad}). Notice that $\vert \hd \vert^2 $ encapsulates 
all the matching scale dependence and 
the bulk of the dependence on the low--scale. 
At the NLO level, the procedure of squaring $\hd$
dictated by perturbation theory, in which terms
of ${\cal O}(\as^2)$ are omitted, gives:
\beq
  \vert \hd \vert^2  =
  \vert C_{7}^{0,\,{\rm eff}} (\mub)\vert^2 
 \left\{   1 + 2\, {\rm Re} 
               \left(\Delta \hd \right)
 \right\}\,,  
\label{dsquare}
\eeq 
where $\Delta \hd$ is defined as (see eq.~(\ref{drewrite})): 
\beq
 \Delta \hd \equiv 
 \f{\hd -C_7^{0,\,{\rm eff}}(\mub)} {C_7^{0,{\rm eff}}(\mub)} =
 \left(\f{\as(\mub)}{4\pi}\right) 
 \f{C_7^{1,{\rm eff}}(\mub) +V(\mub)}{C_7^{0,{\rm eff}}(\mub)} \,.  
\label{correct}
\eeq
If $\vert \Delta \hd\vert $ is not small, the formally 
Next to Next to Leading Order (NNLO)
%NLO (NNLO) 
term $|\Delta \hd|^2$, dropped in eq.~(\ref{dsquare}), is numerically 
relevant. Its omission can lead to 
branching ratios with 
a large dependence on $\mu_b$, or in extreme 
cases, even to negative values for the branching ratio. 
In these situations, the truncation of the perturbative expansion 
of $\hd$ at the NLO level is certainly not justified. 
As we shall see, the size of $\vert \Delta \hd\vert$ 
depends crucially on the values of 
$\xx$, $\yy$ and $\mh$. 
We split our analysis 
as follows. In Sect.~\ref{amplitudexy}, we 
disentangle the effect of the 
couplings $\xx$ and $\yy$ by 
studying the NLO contributions to the Higgs 
components $ \hd_{\smallxy}$ and $ \hd_{\smallyy}$
of the reduced amplitude $\hd$.
In Sect.~\ref{realcouplings} we 
illustrate predictions for the branching ratio for 
real $\xx$ and $\yy$ couplings, giving particular 
emphasis to cases in which the results are highly
unstable or altogether 
unacceptable. In Sect.~\ref{typeoneandtwo}
we discuss in detail Type~II models, for which the NLO 
corrections are under
control, and Type~I models for which the reliability of the 
theoretical prediction depends strongly 
on the point of parameter 
space considered. Finally, in Sect.~\ref{complexcouplings}, we 
study branching ratios and CP asymmetries in the presence of 
complex couplings, and outline 
regions of parameter space where these predictions are 
trustworthy.

%%%%%%%%%%%%%%%%%%%%%%%%%%%%%%%%%%%%%%%%
%
\subsection{Amplitude: $\xx$--$\yy$ independent analysis}
\label{amplitudexy}

\begin{figure}[p]
\begin{center} 
\epsfxsize=12.5 cm
\leavevmode
\epsfbox[130 130 500 670]{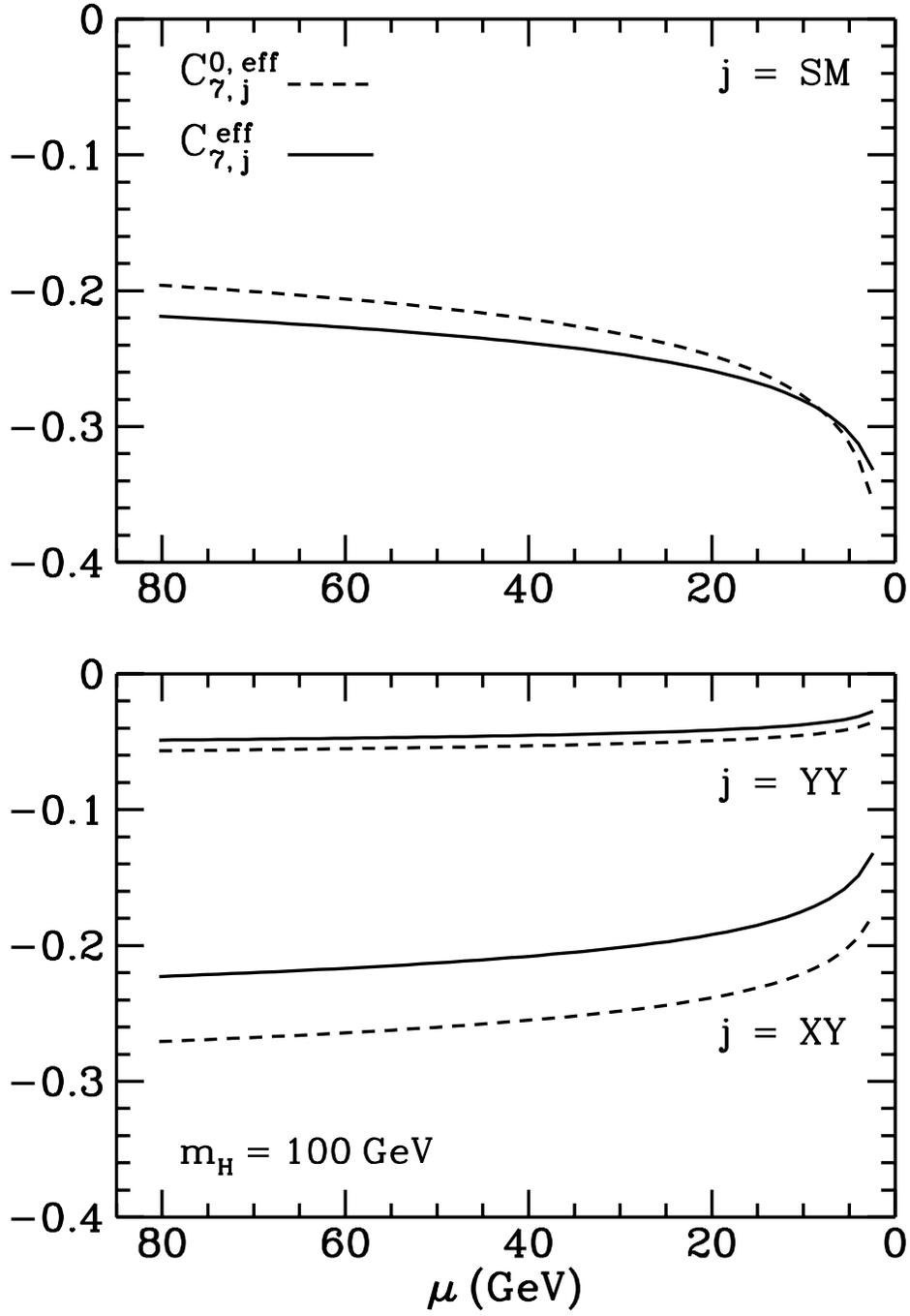}
\end{center}
\caption[f5]{\tenrm{RGE evolution of the Wilson coefficients
 $C_7^{\,{\rm eff}}$, at the LO (with NLO $\as$) (dashed lines)
 and at the NLO (solid lines), for $\muw = \mw$. The upper frame
 shows the SM coefficients, the lower one the coefficients
 $C_{7,\smallyy}^{\,{\rm eff}}$, $C_{7,\smallxy}^{\,{\rm eff}}$,
 for $\mh = 100\,$GeV.  The needed input
 parameters are fixed at their central values listed in
 Table~\ref{inputpar} in Appendix~\ref{InputsScales}.}}
\label{wcoeff}
\end{figure}

\begin{figure}[t]
\begin{center} 
\epsfxsize=12.5 cm
\leavevmode
\epsfbox[130 265 495 535]{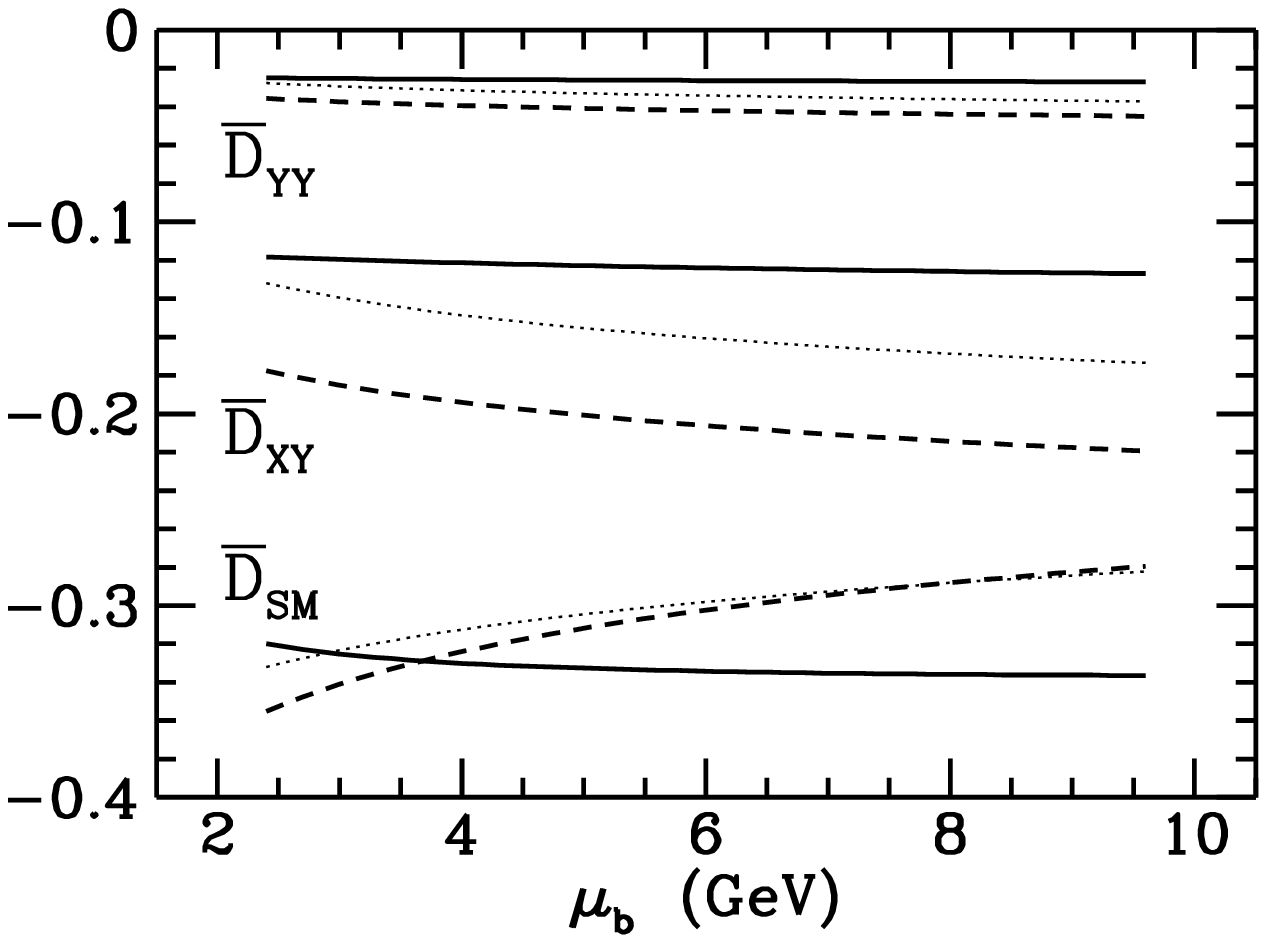}
\end{center}
\caption[f5]{\tenrm{Low--scale dependence of the term $\hd$, (solid
 line), of the LO coefficient $C_7^{0,\,{\rm eff}}(\mub)$ (with
 NLO $\as$) (dashed line), and of $C_7^{\,{\rm eff}}(\mub)$ (dotted
 lines), for $\muw =\mw$. For $\hd_{\smallxy}$ and $\hd_{\smallyy}$,
 the value $\mh = 100\,$GeV was used. The needed input parameters are
 fixed at their central values listed in Table~\ref{inputpar} in 
 Appendix~\ref{InputsScales}.}}
\label{wdterms}
\end{figure}

\noindent 
Before analyzing the size of the various corrections within 2HDMs, it
is worth reviewing the situation in the SM. There, as it is was first
observed in~\cite{CMM}, the NLO contribution to the matching condition
tends to cancel the contribution due to the NLO evolution from the
matching scale $\muw$ to the low--energy scale $\mub$ (see upper 
frame in Fig.~\ref{wcoeff}). 
Indeed, $C_{7,\smallsm}^{1,\,{\rm eff}}$, which is $12\%$ of
$C_{7,\smallsm}^{0,\,{\rm eff}}$ at $\muw$, is practically
negligible for $\mub \in [2.4,9.6]\,$GeV. 
The dominant NLO effect is therefore due to $V_{\smallsm}(\mub)$
which, in the same range of $\mu_b$, is at most $20\%$ of 
the leading term $C_{7,\smallsm}^{0,\,{\rm eff}}(\mub)$ and 
vanishes exactly at $\mub \sim 3.5$\,GeV. 
These features are illustrated in Fig.~\ref{wdterms}. 
The size of $V_{\smallsm}(\mub)$ results from 
cancellations among the individual contributions of
the four operators ${\cal O}_1$, ${\cal O}_2$, ${\cal O}_7$ 
and ${\cal O}_8$ retained in eq.~(\ref{vdefine}).
The details of the above described cancellations 
and, more generally, of the distribution of the complete NLO 
correction among the individual terms in eq.~(\ref{drewrite}), 
are specific to the $\overline{MS}$ scheme (with 
anticommuting $\gamma_5$) and could be altered by a
different choice of scheme. We remind that only the 
complete $\hd_{\smallsm}$ term as well as the 
final $\bbrx$ are scheme--independent, up to higher order. 


\begin{figure}[p]
\begin{center} 
\epsfxsize=12.5 cm
\leavevmode
\epsfbox[130 130  495 665]{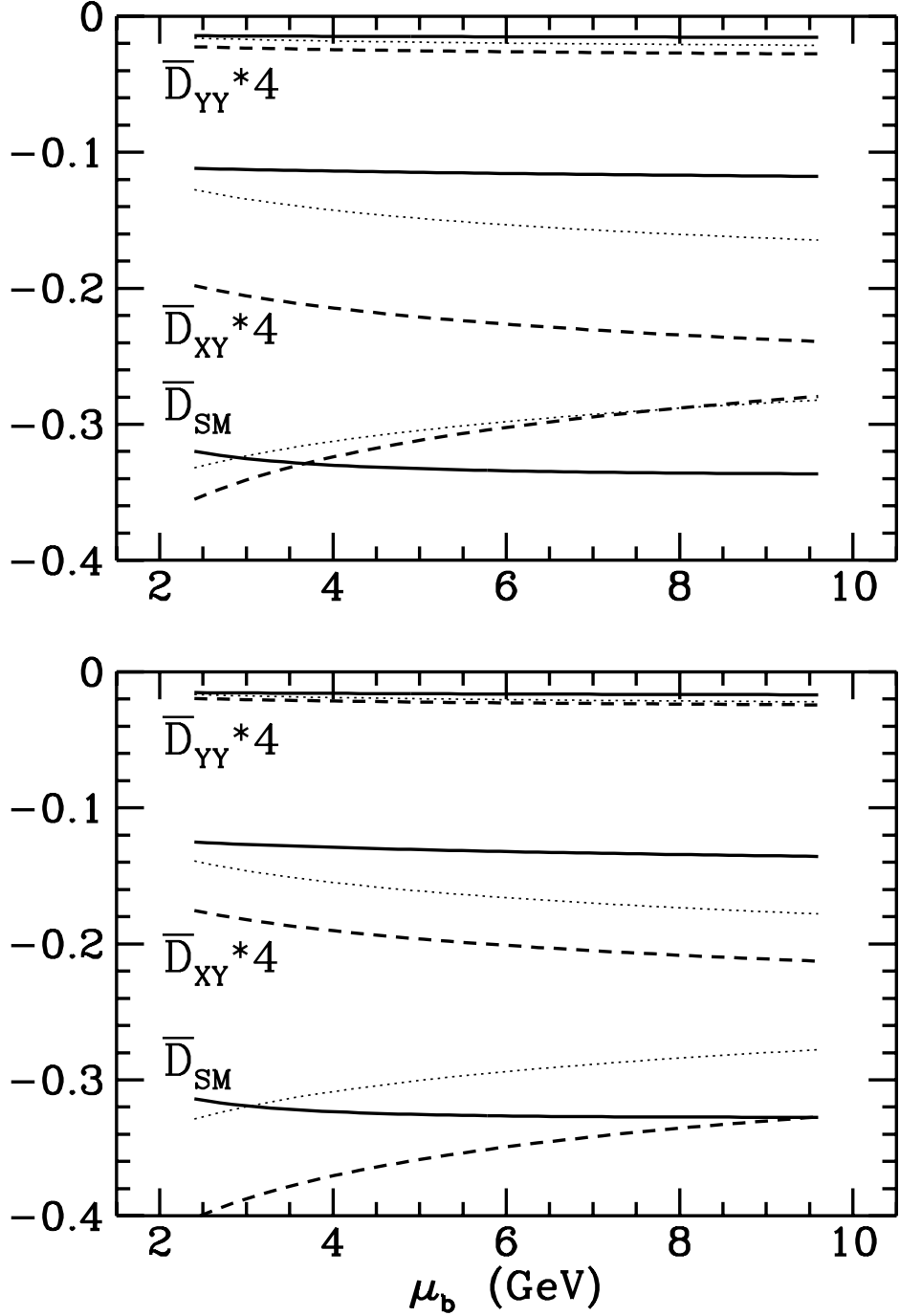}
\end{center}
\caption[f5]{\tenrm{Low--scale dependence of the term $\hd$, (solid
 lines), of the LO coefficient $C_7^{0,\,{\rm eff}}(\mub)$ (with
 NLO $\as$) (dashed lines), and of $C_7^{\,{\rm eff}}(\mub)$ (dotted
 lines), for $\mh=500\,$GeV. The matching scale $\muw=\mw$ is used
 in the upper frame and to $\muw=\mh$ in the lower one. 
 The needed input parameters are fixed at their
 central values listed in Table~\ref{inputpar} in 
 Appendix~\ref{InputsScales}.}}
\label{wdterms500500}
\end{figure}

One is naturally led to ask whether the cancellations among different
sets of NLO corrections observed in the SM are spoiled in 2HDMs.
To address this issue,
we investigate the building blocks $\hd_\smallxy$ and 
$\hd_\smallyy$ defined in eq.~(\ref{decompd}), for 
a specific value of $\mh$. 
In Fig.~\ref{wdterms}, we show the real 
parts of $\hd_{\smallxy}$ and $\hd_{\smallyy}$
for $\mh=100\,$GeV, together with the real part of
$\hd_{\smallsm}$. Even when considering real 
$\xx$ and $\yy$ couplings, imaginary parts to $\hd$ come from 
the absorptive terms of the loop corrections 
in~(\ref{vdefine}).
As it can be seen from eq.~(\ref{dsquare}),
the imaginary parts of $\hd_{\smallxy}$, $\hd_{\smallyy}$
and $\hd_{\smallsm}$ do not contribute to $\bbrx$
at this order of perturbation
theory if $\xx$ and $\yy$ are real. In the remainder of this Section, 
therefore, any reference to these components is 
understood as a reference to their real parts. 

Unlike in the SM, the NLO corrections coming from 
$C_{7,\smallxy}^{1,\,{\rm eff}}(\mub)$ and $V_{\smallxy}(\mub)$,
have roughly the same size ($\sim 20\%$) and the same sign, for
$\mh=100\,$GeV. The combined correction 
\beq
 \Delta \hd_\smallxy \equiv 
 \f{\hd_\smallxy -C_{7,\smallxy}^{0,\,{\rm eff}}(\mub)}
   {C_{7,\smallxy}^{0,\,{\rm eff}}(\mub)}  = 
 \left(\f{\as(\mub)}{4\pi}\right) 
 \f{C_{7,\smallxy}^{1,\,{\rm eff}}(\mub) +V_{\smallxy}(\mub)}
   {C_{7,\smallxy}^{0,\,{\rm eff}}(\mub)}   
\label{correctxy}
\eeq
amounts to the considerable values of $-43\%$ to 
$-36\%$ when varying $\mub \in [2.4,9.6]\,$GeV.  
(We warn here that all components of the LO Wilson coefficient 
$ C_7^{0,\,{\rm eff}}(\mub)$ 
discussed in this Section and plotted in 
Figs.~\ref{wcoeff} and~\ref{wdterms}, are
evaluated with NLO $\as$.)
Differently than in the SM, 
there is no scale $\mub$ in the range considered, 
at which the LO and NLO prediction for $\hd_\smallxy$ coincide.  
Similar results hold for $\hd_\smallyy$.

The change in matching scale from $\muw = \mw$ (value used for
Fig.~\ref{wdterms}) to, say, $\muw = \mh$, is practically 
inconsequential for $\mh=100\,$GeV. It becomes very relevant for 
large values of $\mh$, since it crucially affects the size of the NLO 
corrections. For $\mh=500\,$GeV, as shown in 
Fig.~\ref{wdterms500500}, the correction $ \Delta \hd_\smallxy $
reduces to $-37\%$, $-31\%$ when varying $\mub$ in the usual interval
$[2.4,9.6]\,$GeV, whereas values of $-51\%$, $-45\%$ are obtained for 
$\muw=\mw$. Notice that, for $\mh=500\,$GeV, 
$\vert \hd_{\smallxy} \vert$ is roughly ten times smaller than 
$\vert \hd_{\smallsm} \vert$ and $\hd_{\smallyy}$ completely
negligible. 
When using the same matching scale $\muw=500\,$GeV for the 
SM contribution, the cancellation between the
NLO correction to the matching condition and the NLO evolution of
the Wilson coefficient $C_{7,\smallsm}^{\,{\rm eff}}$ does not 
occur anymore. Large cancellations are instead 
observed between $V_{\smallsm}(\mub)$ and 
$C_{7,\smallsm}^{1,\,{\rm eff}}(\mub)$. The complete 
correction $\vert \Delta \hd_\smallsm  \vert $
ranges still between $19\%$ and $2\%$ for 
$\mub \in [2.4,9.6]\,$GeV, but the point where NLO and LO predictions
for $\hd_{\smallsm} $ coincide is pushed to the higher end of 
$\mub$. These results imply that for heavy
$H^+$ and large enough $\xx$ and $\yy$ couplings (to lift at least 
$\hd_{\smallxy}$ to be of ${\cal O}(\hd_{\smallsm})$), a choice of 
$\muw$ of ${\cal O}(\mh)$, instead than ${\cal O}(\mw)$, minimizes 
the size of the NLO corrections. 

We draw attention to the fact that the sensitivity of the 
reduced amplitude $\hd$ to $\mh$ is weaker than that of the 
Wilson coefficient $C_7^{\,{\rm eff}}(\muw)$ at the matching 
scale. It is interesting to see that the coefficients 
$C_{7,\smallsm}^{\,{\rm eff}}$ and 
$C_{7,\smallxy}^{\,{\rm eff}}$ are almost identical at the matching 
scale $\muw$ for $\mh=100\,$GeV.
This feature is clearly visible 
in Fig.~\ref{wcoeff}, where the matching scale $\muw = \mw$ is 
used. (It remains true for all $\muw \in [\mw,m_t]$.)
We observe in Fig.~\ref{wcoeff} that the RGE flow 
from $\muw$ to $\mu_b$ introduces a large gap between 
$C_{7,\smallsm}^{\,{\rm eff}}(\mub)$ and 
$C_{7,\smallxy}^{\,{\rm eff}}(\mub)$, which is 
then somewhat widened at the level of the reduced amplitude $\hd$ 
by the inclusion of the $V(\mub)$--contribution.
At $\mub \in [2.4,9.6]\,$GeV, $\hd_\smallxy$ is 
about a factor of three smaller than $\hd_\smallsm$. 

When going to physical observables such as $\bbrx$,     
the individual building blocks of $\hd$, 
discussed above in a coupling--independent 
way, are weighted according to the values 
of the couplings $\xx$ and $\yy$.

\begin{figure}[p]
\begin{center} 
\epsfxsize=13.0 cm
\leavevmode
\epsfbox[125 140 500 660]{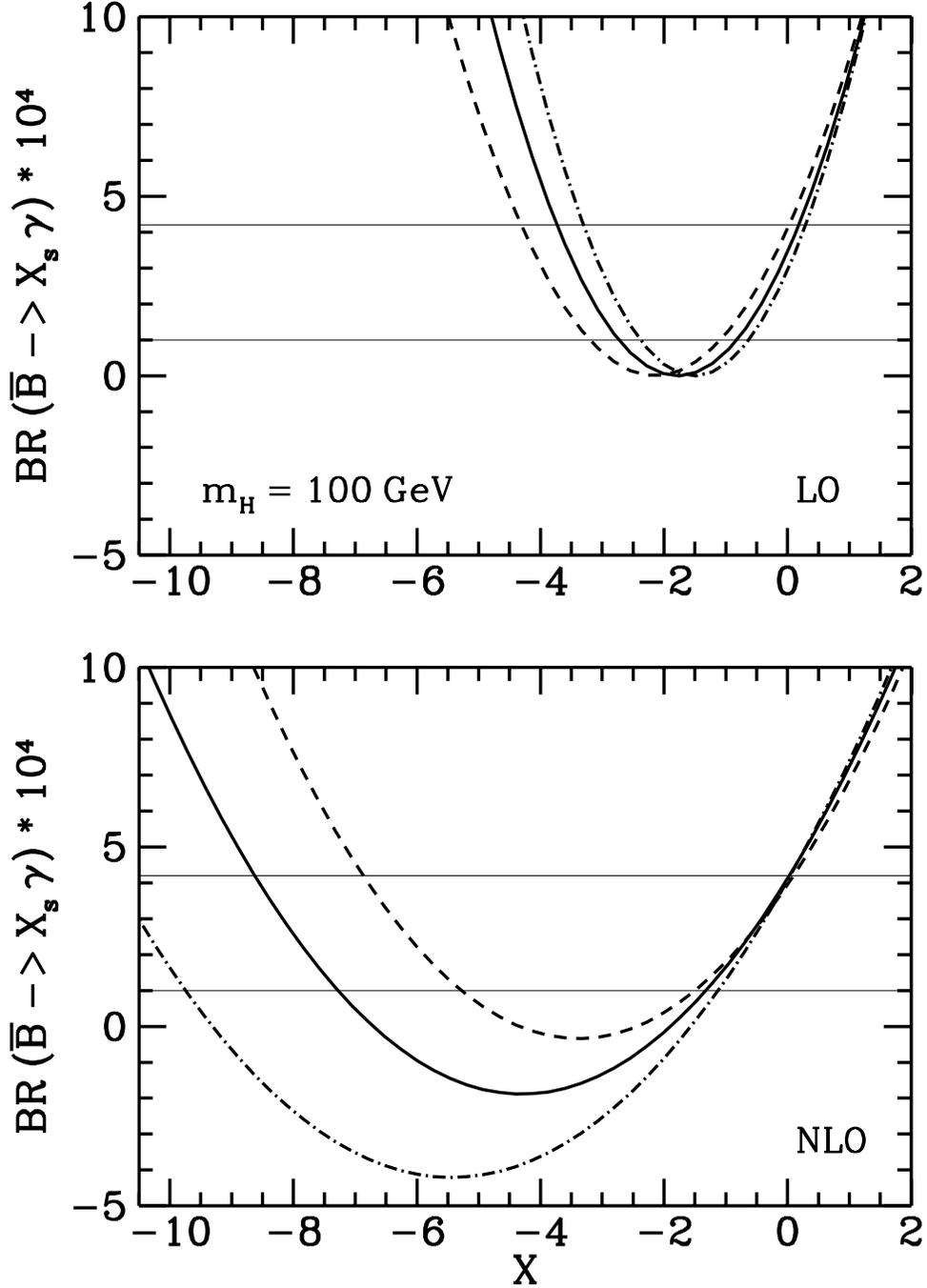}
\end{center}
\caption[f5]{\tenrm{Branching Ratio as a function of the coupling
 $\xx$, for $\yy=1$ and $\mh=100\,$GeV. The upper (lower) frame
 shows the LO (NLO) result for 
 $\mub=4.8\,$GeV (solid lines),
 $\mub=2.4\,$GeV (dashed lines) and
 $\mub=9.6\,$GeV (dash--dotted lines),
 and matching scale $\muw = \mh=100\,$GeV.
 The needed input parameters are fixed at their central values listed
 in Table~\ref{inputpar} in Appendix~\ref{InputsScales}. Superimposed
 is the range of values allowed by the CLEO measurement.}}
\label{figxreal}
\end{figure}

%%%%%%%%%%%%%%%%%%%%%%%%%%%%%%%%%%%%%%%%
%
\subsection{Branching Ratio: Real Couplings}
\label{realcouplings}

\noindent 
The almost perfect flatness of $\hd_{\smallyy}$ and $\hd_{\smallxy}$
shown in Figs.~\ref{wdterms} and~\ref{wdterms500500}, 
should not lead to the conclusion
that the NLO prediction for $\bbrx$ is well--behaved. 
It was explicitly shown in~\cite{BKP}, by expanding the Wilson 
coefficients around $\mub=m_b$, that the dominant 
scale dependence
of the form $\as(m_b) \ln(\mu_b/m_b)$ is cancelled  
in the complete NLO expression for 
$\hd$ in eq.~(\ref{drewrite}) and consequently also in 
$\vert \hd \vert^2$.  
If the NLO corrections are large enough to reduce substantially the
magnitude of the LO term, $\vert \hd \vert^2$ becomes  
sensitive to 
higher order dependence on $\mu_b$ of the form
$\as^2(m_b) \ln^p(\mu_b/m_b)$ ($p=1,2$).
The flatness of $\hd$ seems to indicate that the omitted NNLO term
$\vert \Delta \hd \vert^2$ would cancel to a large extent 
this remaining  $\mu_b$ dependence.
Nevertheless, whether
the omitted $\vert \Delta \hd \vert^2$ in eq.~(\ref{dsquare})
is the bulk of the NNLO
corrections can only be decided when a complete NNLO calculation
is at hand.
 
It is clear that the
reliability of the NLO prediction for the branching ratio,
which is linked to the size of $\Delta \hd$,
depends on the values of $\xx$ and $\yy$.
For a given $\mh$, it is possible to 
choose these couplings in such a way that
the reduced amplitude $\hd$ is dominated, for example, 
by the Higgs contribution 
$\hd_\smallxy$. For such points in the parameter 
space $\{\xx,\yy,\mh\}$, 
the size of the NLO corrections to the branching ratio 
is then roughly $2 \mbox{Re} (\Delta \hd_{\smallxy}) $, i.e.
about $-80\%$ for $\mh=100$ GeV.
In this case, the NLO corrections substantially reduce
the leading order prediction. As expected, the resulting 
scale dependence of the branching ratio $\bbrx$ is large, viz.
about $40\%$. For $\yy=1$ and the same value of $\mh$,  
$\hd_\smallxy$ completely dominates the branching ratio for
$\vert \xx \vert  \ge 20$, outside 
the range shown in Figs.~\ref{figxreal}. 
The prediction for $\bbrx$ is then,
however, far above the band of values allowed by the CLEO
measurement. 

Moreover, it is possible to choose $\xx$ and $\yy$ in such a 
way that $ {\rm Re} ( \Delta \hd ) < -50\%$. 
As eq.~(\ref{dsquare}) shows, 
this choice leads to a negative NLO prediction for $\bbrx$.
Needless to say, in such a situation 
higher order corrections to the NLO calculation are mandatory
to obtain sensible results. In our representative case of 
$\mh =100\,$GeV and $\yy = 1$ illustrated in  
Fig.~\ref{figxreal}, the range of $\xx$ corresponding
to a negative branching ratio is roughly $-10 < \xx < -2$.  
A comparison with the upper
frame of this Figure shows that, given the very large scale dependence
of the LO calculation, one could have guessed the pathological
situation which is encountered at the NLO level. 

\begin{figure}[t]
\begin{center} 
\epsfxsize=12.5 cm
\leavevmode
\epsfbox[135 270 500 530]{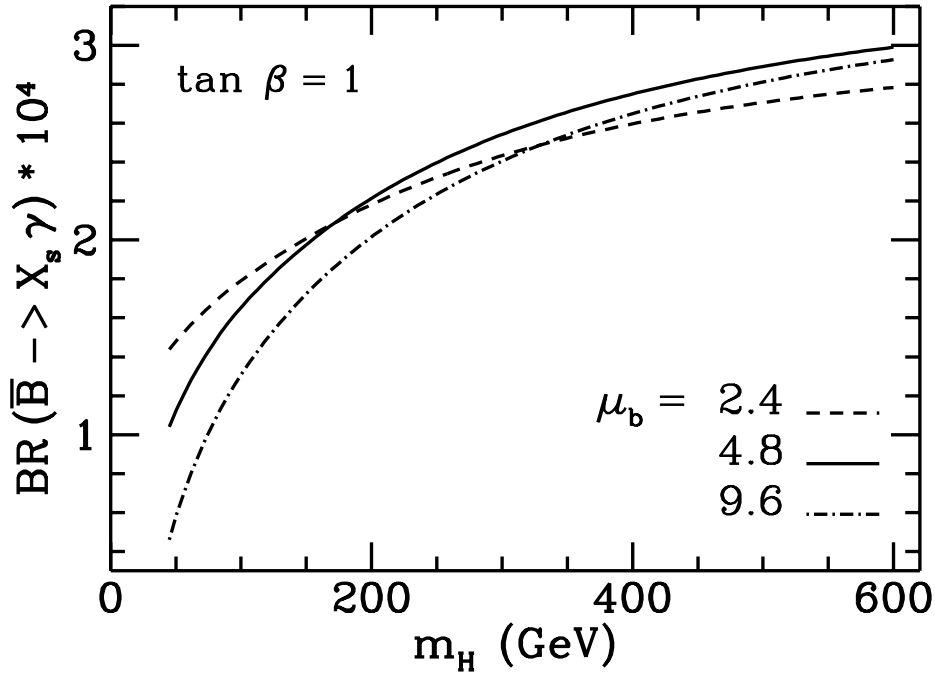}
\end{center}
\caption[f5]{\tenrm{Branching Ratio as a function of $\mh$ 
in a 2HDM of Type I, for three different choices of the low 
scale: $\mu_b = 2.4,4.8,9.6\,$GeV and for $\muw = \mh$.
The needed input parameters are fixed at their
central values listed in Table~\ref{inputpar} in 
Appendix~\ref{InputsScales}.}}
\label{brtypeone}
\end{figure}

Reliable predictions for the branching ratio 
with a mild scale dependence are obtained only
for $\xx>-1$ (see lower frame of Fig.~\ref{figxreal}).
For these values, the SM contribution dominates, but 
the Higgs contribution proportional to $\xys$ is still large enough 
to produce 
a sharp raise of $\bbrx$ when $\xx$ increases from $-1$ to $2$. 
Notice that for $\yy=1$ the values $\xx=-1$ and $\xx=1$ correspond 
respectively 
to the ordinary 2HDM of Type I and Type II with 
$\tanb = 1$. 
Type~I models, however, are not always so stable and 
well--behaved as in the case described above. For $\yy=1$ 
and $\xx=-1$ (corresponding to $\tanb = 1$), it is 
enough to lower $\mh $ to $45\,$GeV to find a low--scale dependence 
of $^{+40\%}_{-60\%}$ (see Fig.~\ref{brtypeone}). 
Keeping $\mh$ at $100\,$GeV and increasing $\yy$ to $2$, 
we find that for $\xx=-2$ (corresponding to a Type~I model with 
$\tanb = 0.5$), the branching ratio is negative for all values of
$\mub \in [2.4,9.6]\,$GeV. 

\begin{figure}[p]
\begin{center} 
\epsfxsize=13.0 cm
\leavevmode
\epsfbox[125 140 500 660]{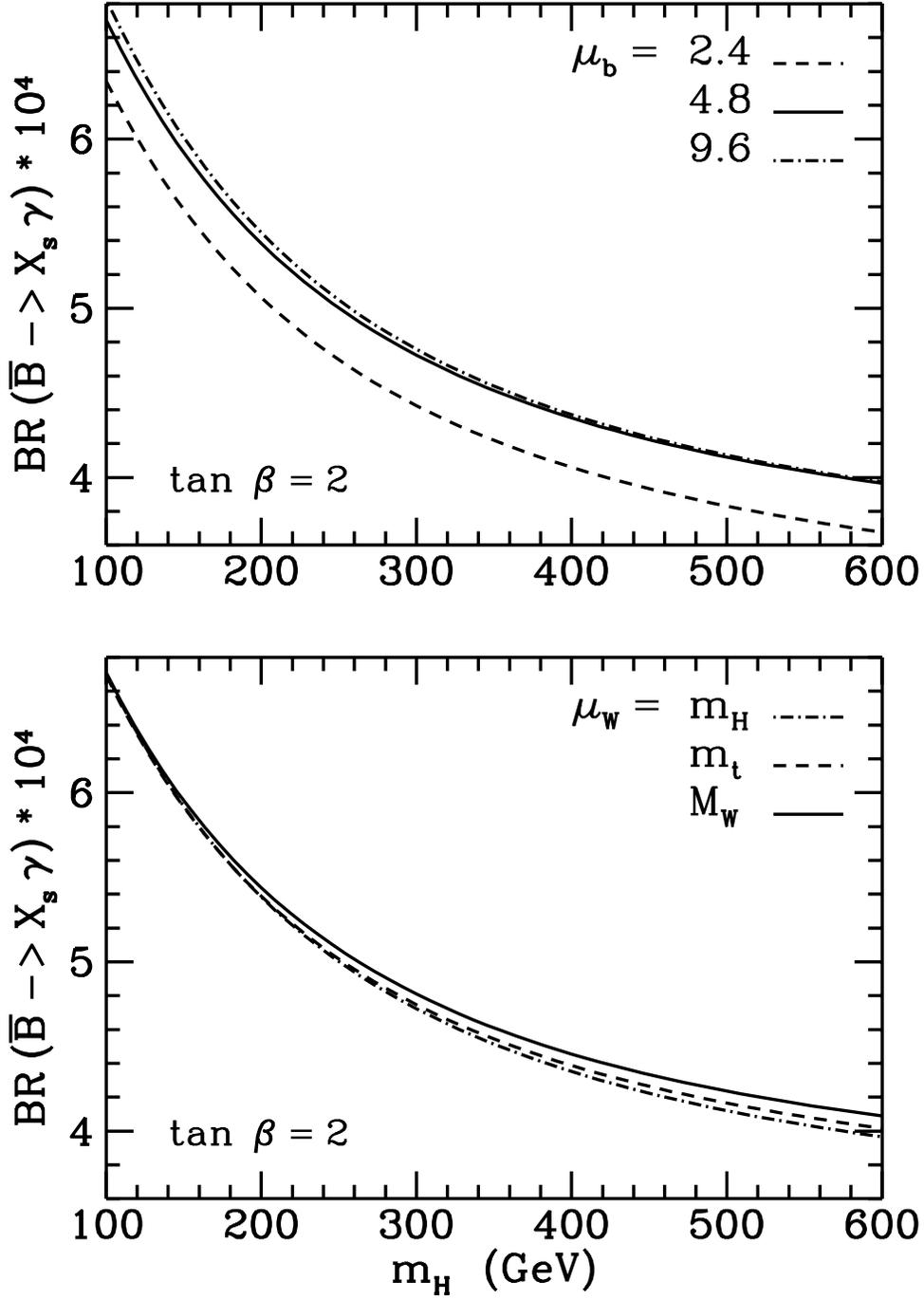}
\end{center}
\caption[f5]{\tenrm{Branching Ratio as a function of $\mh$ 
in a 2HDM of Type II, 
for three different choices of the low--scale:
$\mub = 2.4,4.8,9.6\,$GeV and $\muw = \mh$
(upper frame), and for 
for three different choices of the matching scale:
$\muw = \mw, m_t, \mh$ and $\mub = 4.8\,$GeV.
The needed input parameters are fixed at their
central values listed in Table~\ref{inputpar} in 
Appendix~\ref{InputsScales}.}}
\label{brscales}
\end{figure}

Going to heavier Higgs masses, e.g. $\mh=500\,$GeV, we find for 
$\yy=1$ the same qualitative features as in Fig.~\ref{figxreal}, 
but shifted to larger ranges of $\vert \xx \vert$. We should warn,
however, that for large enough $\vert \xx\vert$, when the 
Higgs contribution dominates over the SM one, the stability 
of the branching ratio becomes worse. The NLO corrections 
have a more dramatic dependence on the particular matching scale 
chosen, than in the case $\mh =100\,$GeV. 
Indeed, for $\muw=\mh=500\,$GeV, the 
relevant correction for the branching ratio
$2 {\rm Re}(\Delta \hd_{\smallxy})$ is $\sim -70\%$, but 
exceeds $-100\%$ for $\muw=\mw$.

\begin{figure}[t]
\begin{center} 
\epsfxsize=13.0 cm
\leavevmode
\epsfbox[125 462 500 715]{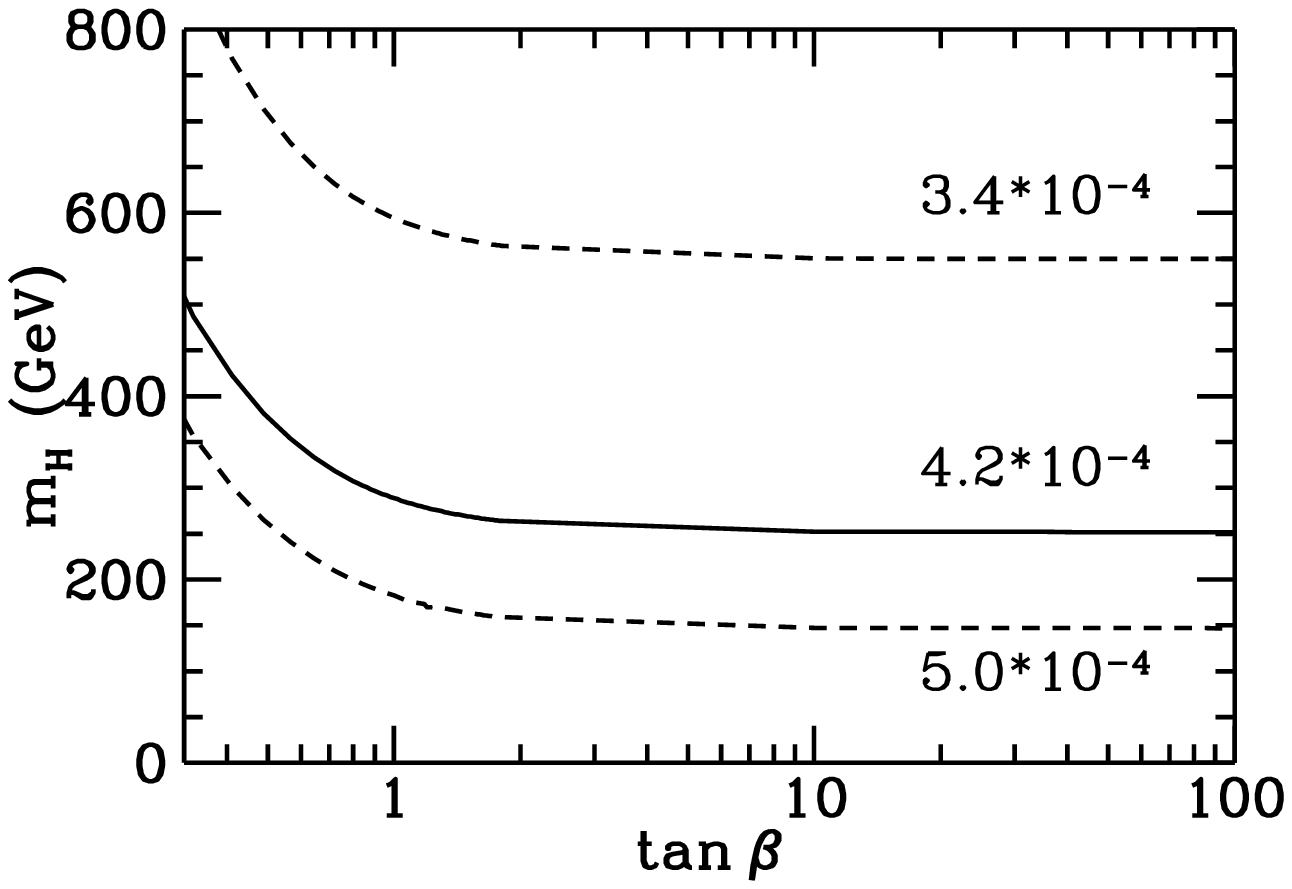}
\end{center}
\caption[f5]{\tenrm{ Contour plot in $(\tan \beta,m_H)$ obtained by 
using the NLO expression for the branching ratio $\bbrx$ 
and possible experimental upper bounds (see text). The allowed region 
is below the corresponding curves.}}
\label{figcontour}
\end{figure}
%

%%%%%%%%%%%%%%%%%%%%%%%%%%%%%%%%%%%%%%%%
%
\subsection{Type I and Type II Models}
\label{typeoneandtwo}

\noindent 
Theoretical predictions for the branching ratio in 
Type~II models stand, in general, on a rather solid ground.
In Fig.~\ref{brscales}, we show these predictions in a 
Type~II model with $\tanb = 2$, for $100\le \mh \le 600\,$GeV. 
This value of $\tanb$ is particularly interesting
since already for $\tanb \ge 2$, the branching 
ratio becomes insensitive to the actual value of this variable:
the contribution $\hd_{\smallyy}$ is, in fact, suppressed by the
coupling $\vert \yy \vert^2 \le 1/4$, whereas the contribution 
$\hd_{\smallxy}$ is multiplied by the coupling $\xys =1$, for any
value of $\tanb$. 
For $\tanb \le 2$, the branching ratio grows very rapidly 
when $\tanb$ decreases
and can be made compatible with existing measurements only 
for large values of $\mh$. 

The upper frame of Fig.~\ref{brscales} shows the low--scale 
dependence of $\bbrx$ for matching scale $\muw = \mh$, for 
$\mh>100\,$GeV. It is less than $10\%$ for any value of 
$\mh$ above the LEP lower bound of $45\,$GeV. Such a small scale 
uncertainty 
is a generic feature of Type~II models and remains true for 
values of $\tanb$ as small as $0.5$.
The lower frame in Fig.~\ref{brscales} shows 
the (very weak) matching scale dependence as obtained by varying
$\muw$ in the interval $[\mw, {\rm Max}(m_t, \mh)]$. We should point 
out that the lowest estimate of $\bbrx$ comes from the largest 
value of $\muw$ in this interval. We 
disagree on this point with
ref.~\cite{CDGG} where 
the lowest estimate is obtained with the smallest value of 
$\muw$. The result of~\cite{CDGG} is presumably due to the 
neglect of the term proportional to $\ln (\muw^2/\mw^2)$ for the
coefficients $C_{1}^{1,\,{\rm eff}}(\muw)$ and  
$C_{4}^{1,\,{\rm eff}}(\muw)$ in eq.~(\ref{coeffNLO}).

In Type~II models, the theoretical estimate of $\bbrx$ can be well 
above the range $(1-4.2)\times 10^{-4}$ indicated by the CLEO
Collaboration as the band of experimentally allowed values.
It is therefore interesting to establish with some 
accuracy which values of $\{ \tanb, \mh\}$ are excluded by
possible measurements of the branching ratio $\bbrx$. 
Our results are given in Fig.~\ref{figcontour},
where we show the contour of the region excluded by the upper bound
obtained by CLEO, $4.2\times 10^{-4}$, as well as for other two  
hypothetical values, $3.4 \times 10^{-4}$ and $5.0 \times 10^{-4}$; 
the latter one is not far from the upper bound obtained by the 
ALEPH Collaboration. These contours are obtained by finding the minimum
of $\bbrx$, when varying simultaneously the 
input parameters within their errors (see 
Table~\ref{inputpar} in Appendix~\ref{InputsScales}) and 
the two scales $\mu_b$ and $\mu_W$ in the ranges 
$2.4\le \mu_b \le 9.6\,$GeV and 
$\mw \le \muw \le {\rm Max}(m_t, \mh)$.
For $\tanb = 0.5,1,5$, we exclude respectively 
$\mh \le 375$,$\,289$,$\,255\,$GeV.
Notice that the flatness of the curves shown in Fig.~\ref{brscales}
towards the higher end of $\mh$, 
causes a high sensitivity of these bounds on all details 
of the calculation. Different treatments of the infrared
singularities when the photon gets soft (e.g. with a cut in the 
photon energy~\cite{CMM}) and the possible expansion of 
$1/\Gamma_{\smallSL}$ in powers of $\as$ could alter the 
branching ratio at the $1\%$ level, i.e. well within the 
estimated theoretical uncertainty. These details, however, 
could still produce shifts of several 
tens of GeV, in either direction, in the lower bounds quoted 
above.    

Branching ratios in Type~I models can be reliably predicted
for $\mh > 100\,$GeV and $\tanb > 1$. 
Theoretical results for this range of parameters 
fall within the CLEO band 
$(1-4.2)\times10^{-4}$, as it can be seen in Fig.~\ref{brtypeone}
for $\tanb =1$ and $100 \le \mh \le 600 \,$GeV. Larger values of 
$\tanb$ decrease the Higgs contribution to the branching ratio, 
giving therefore values closer to the SM prediction. Lower values of
$\mh$ 
(and/or $\tanb <1$) can produce results outside the CLEO range. 
For these parameters, however, the theoretical predictions
are unstable under scale variation and, at times, ill--defined. 
As shown in Fig.~\ref{brtypeone}, a scale dependence of 
$^{+40\%}_{-60\%}$ is obtained for $\mh=45\,$GeV and 
$\tanb = 1$. For $\tanb < 1$, negative values of 
%$\bbrx$ 
the branching ratio are found already for $\mh = 100\,$GeV. 

\begin{figure}[p]
\begin{center} 
\epsfxsize=13.0 cm
\leavevmode
\epsfbox[125 135 495 655]{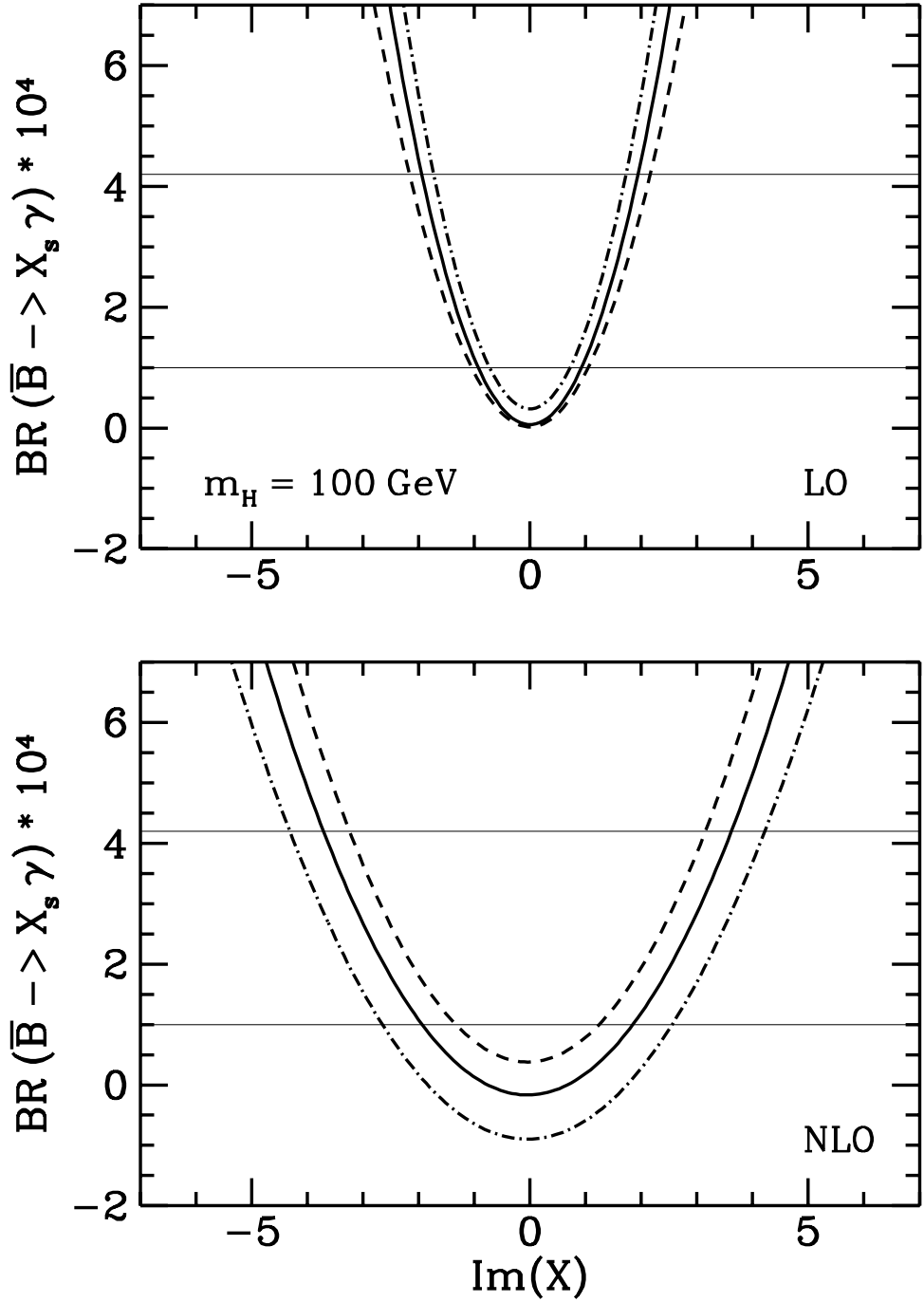}
%\epsfbox[108 179 512 724]{ximag.ps}
\end{center}
\caption[f5]{\tenrm{Branching Ratio as a function of $\imx$, 
 with fixed $\rex=-2$ and $\yy=1$, 
 for $\muw = \mh$ and different low--scales:
 $\mub=4.8\,$GeV (solid lines),
 $\mub=2.4\,$GeV (dashed lines) and
 $\mub=9.6\,$GeV (dash--dotted lines).
 The needed input parameters are fixed at their central values listed
 in Table~\ref{inputpar} in Appendix~\ref{InputsScales}. Superimposed
 is the range of values allowed by the CLEO measurement.
 The upper (lower) frame shows 
 the LO (NLO) result.}}
\label{figximag}
\end{figure}

%%%%%%%%%%%%%%%%%%%%%%%%%%%%%%%%%%%%%%%%
%
\subsection{Branching Ratio: Complex Couplings and Rate Asymmetries}
\label{complexcouplings}

\noindent 
In this section we study several aspects of the decay $\bbxsg$
in the presence of complex $\xx$ and $\yy$ couplings. 
Particularly motivating is the
observation that the measurement of $\bbrx$ yields 
strong constraints on $\ixys$~\cite{KP,GNR}. In 
turn, these constraints limit the possibility of having large 
indirect CP asymmetries in neutral $B$ decays such as 
$B \to \psi K_{{\scriptscriptstyle S}}$~\cite{GNR}. This observation 
is based on
a LO analysis whose reliability and stability under scale
variation was not enquired. 
We plan to investigate this aspect 
and to check how the bound in~\cite{GNR} may be modified by 
the inclusion of NLO corrections. 

The bound of ref.~\cite{GNR} on $\ixys$ is obtained as follows.
Since the LO branching ratio is proportional
to $\vert C_{7}^{0,\,{\rm eff}} (\mu_b) \vert^2 $, the 
upper bound $\bbrx \le 4.2 \times 10^{-4}$ from 
the CLEO measurement 
implies an upper bound on  
$\vert C_{7}^{0,\,{\rm eff}} (\mu_b) \vert^2 $, and therefore on 
$\ixys$ when:
\bea
 {C_{7}^{0,\,{\rm eff}} (\mu_b)} & = & 
 \left\{
  {C_{7,\smallsm}^{0,\,{\rm eff}} (\mu_b)} +
 \vert \yy \vert^2
  {C_{7,\smallyy}^{0,\,{\rm eff}} (\mu_b)} +  
 \rxys 
    {C_{7,\smallxy}^{0,\,{\rm eff}} (\mu_b)}
 \right\}  + i \,
 \ixys 
    {C_{7,\smallxy}^{0,\,{\rm eff}} (\mu_b)}   \nn \\ 
                                 & = &  
 i \, \ixys 
    {C_{7,\smallxy}^{0,\,{\rm eff}} (\mu_b)} \,,   
\label{imagcondition}
\eea
i.e., when the real parts of the charged Higgs contributions
cancel the SM coefficient 
$C_{7,\smallsm}^{0,\,{\rm eff}} (\mu_b)$.
An inspection of the upper frame of Fig.~\ref{figxreal}
shows that, for $\mh=100\,$GeV and $\yy=1$,  
a vanishing branching ratio is induced by the 
the real coupling $\xx=-2$.
The choice of complex couplings $\xx$ and $\yy$ with 
$\yy =1 $ and $\rex = -2 $, 
therefore, fulfills to a good approximation the 
condition~(\ref{imagcondition}), for all values of $\imx$.
The corresponding branching ratio, 
obtained with the central value of the 
input parameters, $\muw =\mh$, and $\mub=m_b$, and
shown in Fig.~\ref{figximag} as a function of $\imx$, equals 
$4.2\times 10^{-4}$ at $\vert \imx\vert \sim 2$. 
For the chosen Higgs mass, $\mh=100\,$GeV, therefore, the upper
bound on $\vert \ixys \vert $ is $\sim 2$. 
Notice that this procedure yields only a first estimate
for the actual LO bound, since the errors of the input 
parameters have not been considered.

We observe that a variation of the low--scale $\mub$ in the usual 
range leads to large uncertainties for the branching ratio,
throughout the whole range of $\imx$. In particular, 
for $\vert \imx \vert \sim 2$, this uncertainty amounts to 
$\pm 25\%$. When including NLO corrections, the situation
does not improve, as the lower frame of Fig.~\ref{figximag}
shows. It is interesting to see that the value of $\bbrx$
for $\mub=m_b$, at $\vert \imx\vert \sim 2$ drops from 
$4.2\times 10^{-4}$ to roughly  $1\times 10^{-4}$. The 
intersection of the NLO curve for $\mub=m_b$ with the 
CLEO upper bound is at $\imx \sim 4$. This procedure is 
essentially the construction of the NLO bound for 
$\vert \ixys \vert$ in the sense specified before.
(In the actual construction, one should have cancelled also 
small real parts of $C_7^{0\,,{\rm eff}}$ with NLO 
$\as$ and small terms coming from absorptive parts of
virtual corrections.)

The inclusion of NLO corrections verifies explicitly  
the instability of the LO upper bound on $\ixys$. Given the 
fact that the NLO predictions for $\bbrx$ 
are plagued by even larger scale uncertainties, 
it is hard to believe that the NLO candidate qualifies as a 
reliable bound. 

\begin{figure}[p]
\begin{center} 
\epsfxsize=12.5 cm
\leavevmode
\epsfbox[125 183 500 715]{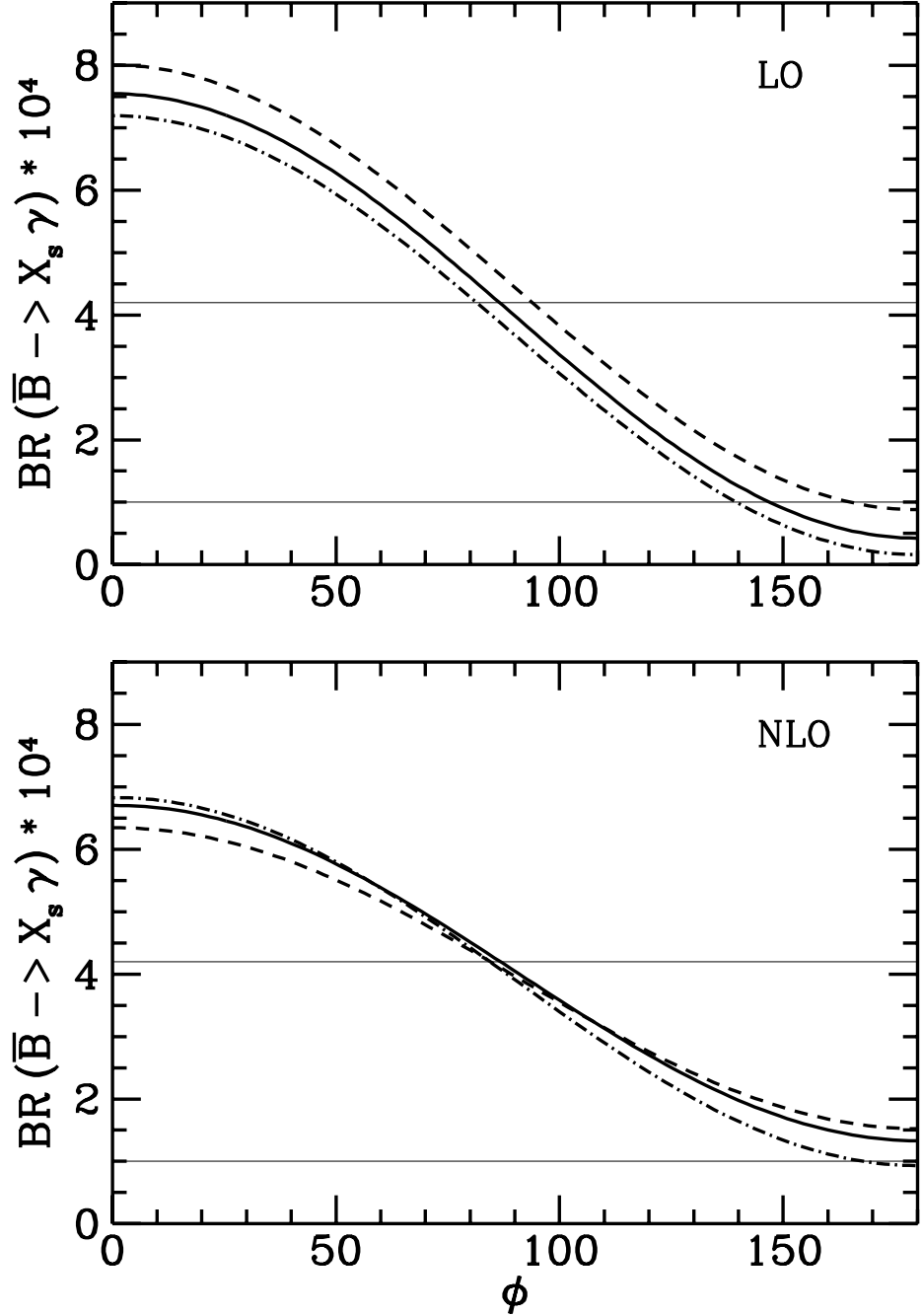}
\end{center}
\caption[f5]{\tenrm{Branching Ratio as a function of $\phi$,
 where $\phi$ parametrizes $\xx$: $\xx= 2 \exp (i\phi)$, for
 $\yy=1/2$, $\mh=100\,$GeV, 
 and $\muw = \mh$. Solid, dashed and dash--dotted lines correspond 
 respectively to $\mub=4.8, 2.4$ and $9.6\,$GeV.
 The needed input parameters are fixed at their central values listed
 in Table~\ref{inputpar} in Appendix~\ref{InputsScales}. Superimposed
 is the range of values allowed by the CLEO measurement.
 The upper (lower) frame shows the LO (NLO) result. }}
\label{figxfi}
\end{figure}

Not all complex $\xx$ and $\yy$ couplings
yield NLO predictions as problematic as those 
shown in Fig.~\ref{figximag}. A typical case in which the 
perturbative expansion of $\bbrx$ 
can be safely truncated at the NLO level is 
identified by: 
$\yy=1/2$, $\xx = 2 \exp (i\phi)$,  and $\mh=100\,$GeV. 
Indeed, for these parameters, the real and imaginary 
part of $\vert \yy \vert ^2 \hd_{\smallyy}$ and 
$ (\xys) \hd_{\smallxy}$ are dominated by $\hd_{\smallsm}$. 
Therefore, the low--scale dependence of the 
branching ratio is not much larger than the very mild one 
obtained for the SM estimate, as shown in Fig.~\ref{figxfi}.
This case is particularly interesting since it gives rise to a
theoretical prediction of $\bbrx$ consistent with the 
CLEO measurement, even for a relatively small value of $\mh$. 
Such a light charged Higgs can contribute to 
the decays of the $t$-quark, through the mode $t \to H^+ b$.

\begin{figure}[p]
\begin{center} 
\epsfxsize=12.5 cm
\leavevmode
\epsfbox[130 140 500 665]{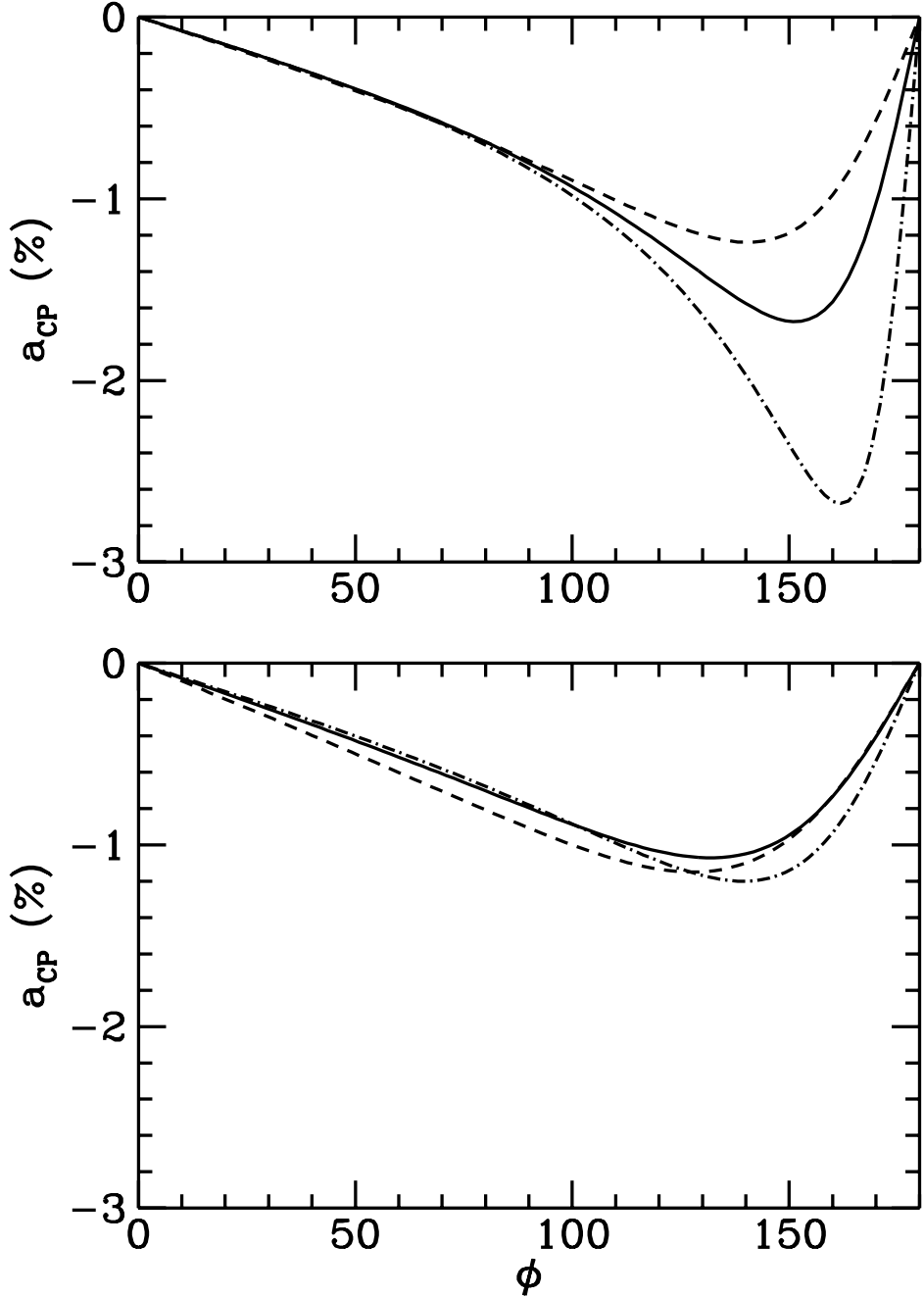}
\end{center}
\caption[f5]{\tenrm{CP rate asymmetry $a_{cp}$ as a function 
 of $\phi$, where $\phi$ parametrizes $\xx$:
 $\xx= 2 \exp (i\phi)$, for $\yy=1/2$, $\mh =100\,$GeV,
 and $\muw = \mh$.
 Solid, dashed and dash--dotted lines correspond respectively to 
 $\mub=4.8, 2.4$ and $9.6\,$GeV.
 The needed input parameters are fixed at their central values listed
 in Table~\ref{inputpar} in Appendix~\ref{InputsScales}. 
 In the upper (lower) frame the LO (NLO) expression for the denominator
 in eq.~(\ref{acpdef}) is used. See text for more details.  
}}
\label{figxcp}
\end{figure}

We now investigate CP asymmetries induced by complex couplings. 
It is well known that in the SM a non--vanishing direct CP rate
asymmetry
\beq
\label{acpdef}
 a_\smallcp = \frac{\bbrx -\barbbrx}{\bbrx +\barbbrx}
\eeq
is due to 
non--trivial relative weak-- and relative strong phases in the 
decay amplitude for $\bsg$ (as well as the 
one for $b \to s \gamma g$). 
We write the amplitude of $\bsg$ and of the CP conjugated 
process as:
\bea
\label{adecomp} 
{\cal A}(\bsg) &=&
(\llu)\, A_u \,  + (\llc)\, A_c \  + (\llt)\, A_t \,;
%  \quad \quad \l_i = V_{ib}V_{is}^*  
       \nonumber \\
{\cal A}(\overline{b} \to \overline{s} \gamma) &=&
(\llu)^* A_u + (\llc)^* A_c + (\llt)^* A_t \,,
\eea 
where the dependence on the CKM matrix is manifest. 
Working in the LO approximation, the quantities $A_u$, $A_c$, and
$A_t$ are real, hence 
$\vert {\cal A}(\overline{b} \to \overline{s} \gamma) 
\vert^2=\vert {\cal A}(b \to s \gamma) \vert^2$, 
and consequently $a_\smallcp =0$. Relative strong phases among $A_u$,
$A_c$, and $A_t$ only appear at the NLO level, due to absorptive terms
in the loop diagrams. Thus, a non--vanishing CP asymmetry results if
$\llu$, $\llc$, and $\llt$ have also relative phases.  In the
approximation $\llu=0$, which we used so far, the CP asymmetry
vanishes, since unitarity implies that $\llc=-\llt$.  Undoing this
approximation, the SM CP rate asymmetry turns out to be 
below $1\%$~\cite{AAG}.

Since the CP rate asymmetry is so tiny in the SM, it 
would be exciting if the imaginary parts of the $\xx$ and 
$\yy$ couplings would induce --together with the absorptive parts
of the NLO loop-functions--
measurable CP rate asymmetries.
In order to investigate this question, we switch off the SM asymmetry
by working in the limit $\llu=0$. 
As an illustrative example we calculate the CP asymmetries for the
same values of parameters as in Fig.~\ref{figxfi}, i.e.,
for $\xx = 2 \exp(i\phi)$, $\yy=1/2$ and for $\mh=100$ GeV,  
where the branching ratio appears to be well--behaved.  
As mentioned above, the NLO prediction for the branching ratio
in the numerator of eq.~(\ref{acpdef})
is required to obtain the first non--vanishing term 
for the rate asymmetry. For the denominator, 
one can either use the LO or the NLO estimate for the two 
branching ratios. The difference amounts to 
higher order terms which are not systematically calculated.
The respective results are shown in the upper and lower frame in 
Fig.~\ref{figxcp} as a function of $\phi$. 
For this specific choice of $\xx$ and $\yy$ the asymmetries 
are rather modest, at the $1\%$ level. The scale dependence 
shown by the lower frame is certainly smaller than that in 
the upper frame. However, since 
both procedure are in principle viable, one cannot conclude that
the result for the CP asymmetries is as reliable as indicated
in the lower frame. 
Although we have not systematically scanned the
parameter space $\{\xx,\yy,\mh\}$, this is our generic result: 
choices of couplings $\xx$ and $\yy$ which render the 
branching ratio stable, i.e. 
not plagued by a large dependence on the scale $\mub$, induce 
small values of $a_\smallcp$.

%%%%%%%%%%%%%%%%%%%%%%%%%%%%%%%%%%%%%%%%
%
\section{Conclusions}
\label{conclusions}

\noindent 
We have presented NLO predictions for the decay 
$\bbxsg$ in generic 2HDMs with flavour--conserving tree--level
neutral couplings. We include in this definition Multi--Higgs Doublet
models where all charged Higgs bosons, except one, are heavy and
completely decoupled at the electroweak scale. Their existence leaves
an imprint only in the Yukawa potential where the couplings $\xx$ and
$\yy$ multiplying respectively the down-- and up--term are not
necessarily correlated, and in general complex. This generalization
allows a simultaneous study of $\bbxsg$ for different types of 2HDMs
by continuously varying the couplings $X$ and $Y$. Results for the
well--known Type~I and Type~II models are then obtained for specific
choices of real couplings: only one combination of $\xx$ and 
$\yy$, usually denoted as $\tanb$, occurs.  
Since supersymmetric models have an enlarged Higgs sector with two
Higgs doublets of Type~II, the results presented here are also a first
step towards a complete evaluation of the rate of $\bbxsg$ in these
models. They constitute already a good approximation for those 
supersymmetric scenarios, such as gauge--mediated supersymmetric 
models with large $\tanb$, 
where the Higgs contribution dominates by far over the 
genuinely supersymmetric contributions~\cite{MMM}. 

Our calculation is carried out using the effective Hamiltonian 
formalism with on--shell operators. 
The NLO matching condition for the decay $\bbxsg$ was already
completely calculated in~\cite{CDGG} and partially in~\cite{CRS}. For
the Wilson coefficients $C_7^{\,{\rm eff}}(\muw)$ and 
$C_8^{\,{\rm eff}}(\muw)$, we find agreement with the existing 
results, but we disagree with ref.~\cite{BKP,CDGG} for the 
matching scale dependence of the 
coefficients $C_1^{\,{\rm eff}}(\muw)$ and $C_4^{\,{\rm eff}}(\muw)$. 
We generalize the solution of the RGE needed to obtain the coefficient 
$C_7^{\,{\rm eff}}(\mu_b)$ at the low--scale $\mu_b$ for 
values of $\muw$ different from $\mw$. 
We also correct the dependence on the Wilson coefficients of one
of the non--perturbative contribution to $\Gamma(\bbxsg)$, which 
is erroneously reported in~\cite{CDGG}. 

We have given predictions for $\bbrx$ in 2HDMs as functions of the 
parameters $\{X,Y,\mh\}$. We found that the theoretical
uncertainties of these NLO calculations are in general larger than
those obtained in the SM. 
The quality of our predictions, therefore, depends strongly on 
the values of the parameters considered. 
As a first step, we summarize the theoretical features of our 
results, ignoring a comparison with the existing experimental 
data. We distinguish several cases:
{\it i}) When these parameters are such that the SM contribution is 
much larger than the Higgs contribution to $\bbrx$, the 
reliability of our predictions does not differ much from that 
in the SM. This, however, does not prevent the $\bbrx$ from 
being rather sensitive to $\mh$. 
{\it ii}) For values of parameters which bring the Higgs 
contributions to the same level as the SM one, with constructive 
interference between the two, we find in general larger scale 
dependence than in the SM.
{\it iii}) When the Higgs contributions are still of the same order as 
the SM contribution, but interfere destructively with it, we find 
very large scale dependences and, in specific cases, we obtain
negative values of $\bbrx$. 
{\it iv}) When $\xx$ and $\yy$ make the Higgs contributions 
largely dominate
over the SM one, we find a scale dependence in $\bbrx$ of $\sim 40\%$
for $\mh$ of ${\cal O}(\mw)$. For much larger values of $\mh$, the 
quality of the theoretical prediction depends strongly on the matching
scale considered. 

Type~II models are typical of case {\it i}) and barely become 
of case {\it ii}) for $\mh$
at the LEP lower bound of $45\,$GeV~\cite{BD} and for 
$\tanb \sim 0.5$. The theoretical uncertainty for such models is 
in general below $10\%$. 
Similar uncertainties are obtained 
for Type~I models when $\mh$ and/or $\tanb$ are large enough. 
Rather unstable results are, however, obtained 
for $\mh$ at the lower end of allowed values and 
$\tanb \ltap {\cal O}(1)$. 

When these problematic regions are avoided, then all predictions 
obtained for Type~I models are consistent with the CLEO 
measurement of $\bbrx$.  The same is not true for Type~II 
models where the 
theoretical predictions are always above the SM result.  
Measurements become then highly constraining. 
Taking into account only the CLEO result, without combining it with the
still preliminary one from ALEPH, we  
exclude $\mh \le 375\,$GeV for $\tanb = 0.5 $ and 
$\mh \le 255\,$GeV for $\tanb = 5$.  
These bounds are very sensitive to 
details in the definition of the branching ratio
(e.g. whether it requires or not an expansion in $\as$ of
$1/\Gamma_{\smallSL}$), which give rise to uncertainties 
usually not included in 
the theoretical error for this observable. 
For the generic 2HDMs which we consider, 
we find wide regions of parameter
space where the theoretical predictions for $\bbrx$ are reliable 
and within the band of values $(1.0-4.2)\times 10^{-4}$ allowed 
by the CLEO measurement. In particular, we find that 
within these models, charged Higgs bosons 
can still be light enough to be produced through a decay of 
the $t$--quark.

Since $\xx$ and $\yy$ are in general complex, new CP violating
effects are induced. In particular, the combination $\ixys$ is 
important as it may 
affect the indirect CP asymmetries in $B\to \psi K_s$ already 
at the LO. It is known that the measurement of $\bbrx$ constrains
$\ixys$~\cite{KP} and an upper bound for this quantity has been 
obtained in~\cite{GNR} using the LO calculation. 
We find that the LO estimate of the branching ratio at the values of
$\xx$ and $\yy$ which determine the upper bound extracted
in~\cite{GNR} is unstable under scale variation. 
The addition of NLO corrections tend to shift the LO bound to 
a higher value. 
Nevertheless, the scale dependence of $\bbrx$ for the new
combination of $\xx$ and $\yy$ needed, 
is still too large to conclude that the bound obtained at the 
NLO level is stable against higher order corrections.
The large instabilities encountered even at this order are 
due to the fact that 
the construction of this upper bound requires the almost
complete cancellation of the SM contribution.

At the NLO, complex values of $\xx$ and $\yy$ induce also direct CP
rate asymmetries, in $\bbxsg$, which is essentially vanishing in the
SM. If sizable, a measurement of this observable, could provide a
handle to detect some of these models. Unfortunately, we find that
combinations of couplings where $\bbrx$ is reliably predicted, lead to
rate asymmetries only at the $1\%$ level.

We find that within 2HDMs, the truncation of the perturbative series
at the NLO level is often inappropriate. This is in sharp contrast
with the SM case, where the large LO theoretical uncertainties are
drastically reduced in the NLO calculation, and the overall size of
the NLO corrections is, in comparison, rather modest. It is somehow
disturbing to find the problematic features described in this paper in
models which structurally do not differ very much from the SM and it
is conceivable that other extensions may suffer from the same
diseases.  One may wonder about the deeper reasons for the
cancellations observed in the SM. Their understanding may help in 
clarifying also the reasons for the features encountered in 2HDMs.

\appendix{}

\section{Wilson Coefficients at $\muw$} %the Matching Scale}
\label{defineCoeff}
\noindent
We list here the functions introduced in the text,
which define the Wilson coefficients at the matching scale
(see eqs.~(\ref{coeffLO})--(\ref{coeffNLOa})).  

\noindent
{\bf SM case}

\noindent
The LO functions are ($x=m_t^2/\mw^2$):
\bea
 C_{7,\smallsm}^0  & = & \frac{x}{24} \, \left[
 \frac{-8x^3+3x^2+12x-7+(18x^2-12x) \ln x}{(x-1)^4} 
                                         \right]      \nn \\[1.5ex]
 C_{8,\smallsm}^0  & = & \frac{x}{8} \, \left[
 \frac{-x^3+6x^2-3x-2-6x \ln x}{(x-1)^4} \right] \,.
\label{wclosm}                                            \\[1.7ex]
\fudge{In the ${\overline{MS}}$ scheme, we have at the NLO:}\\[1.1ex]
 E_0 \quad \,      & = &
 \frac{x (x^2+11x-18)}{12 (x-1)^3}
+\frac{x^2 (4 x^2-16x+15)}{6(x-1)^4} \ln x-\frac{2}{3} \ln x
-\frac{2}{3}                                          \nn \\[1.5ex]
 W_{7,\smallsm}    & = &
 \f{-16 x^4 -122 x^3 + 80 x^2 -  8 x}{9 (x-1)^4} \,
  {\rm Li}_2 \left(\!1\!-\!\f{1}{x} \right)
+\f{6 x^4+46 x^3-28 x^2}{3 (x-1)^5} \ln^2 x           \nn \\
                   &   &
+\f{-102x^5-588 x^4-2262 x^3+3244 x^2-1364 x+208}
    {81(x-1)^5} \ln x                                 \nn \\
                   &   &
+\f{1646x^4+12205x^3-10740x^2+2509x-436}{486(x-1)^4}  \nn \\[1.5ex]
 W_{8,\smallsm}    & = &
 \f{-4 x^4 +40 x^3 + 41 x^2 + x}{6 (x-1)^4} \,
  {\rm Li}_2 \left(\!1\!-\!\f{1}{x} \right)
+\f{ -17 x^3 - 31 x^2}{2 (x-1)^5} \ln^2 x             \nn \\
                   &   &
+\f{-210 x^5+1086 x^4+4893 x^3+2857 x^2-1994 x+280}
     {216 (x-1)^5} \ln x                              \nn \\
                   &   &
+\f{737x^4-14102 x^3-28209x^2+610 x-508}{1296(x-1)^4} \nn \\[1.7ex]
 M_{7,\smallsm}   & = &
 \frac{82x^5\!+\!301x^4\!+\!703x^3\!-\!2197x^2\!+\!1319x\!-\!208-
      \! (162x^4\!+\!1242x^3\!-\!756x^2)\ln x}
        {81 (x-1)^5}                                  \nn  \\[1.5ex]
 M_{8,\smallsm}   & = &
\frac{77x^5-475x^4-1111x^3+607x^2+1042x-140+
      \! (918x^3+1674x^2) \,\ln x}
        {108(x-1)^5}                                  \nn \\[1.7ex]
 T_{7,\smallsm}    &=&  \frac{x}{3} \, \left[
\frac{47x^3-63x^2+9x+7-(18x^3+30x^2-24x)
                       \ln x}{(x-1)^5} \right]        \nn \\[1.5ex]
 T_{8,\smallsm}    &=&  2x \, \left[\frac{-x^3-9x^2+9x+1+(6x^2+6x)
                       \ln x}{(x-1)^5} \right] \,.
\label{wcnlosm}  
\eea

\noindent
{\bf 2HDM case, coupling $\vert Y\vert^2$}

\noindent
The functions relative to charged Higgs contributions, with
coupling $\vert Y\vert^2$, are at the LO ($y=m_t^2/\mh^2$):
\bea
 C_{7,\smallyy}^0  & = &
\frac{1}{3} \, C_{7,\smallsm}^0(x \to y)              \nn \\[1.5ex]
 C_{8,\smallyy}^0  & = &
 \frac{1}{3} \, C_{8,\smallsm}^0(x \to y) \,;            
\label{wcloyy}                                            \\[1.7ex]
\fudge{at the NLO:}                                       \\[1.1ex]
 E_H \quad \,     & = & \frac{1}{36} \, y \, \left[
  \frac{7y^3-36y^2+45y-16+(18y-12) \ln y}{(y - 1)^4} \right]
                                                       \nn \\[1.5ex]
 W_{7,\smallyy}   & = &
 \frac{2}{9}\,y\left[
 \frac{8y^3-37y^2+18y}{(y -1)^4}\,
  {\rm Li}_2 \left(\!1\! -\! \f{1}{y} \right)
+\frac{3y^3+23y^2-14y}{(y -1)^5}\ln^2y \right.         \nn \\
                  &   &
\phantom{ \frac{2}{9}\,y\left[\right.}
+\frac{21y^4-192y^3-174y^2+251y-50}{9(y-1)^5} \ln y    \nn \\
                  &   &
\phantom{ \frac{2}{9}\,y\left[\right.}
\left .
+\frac{-1202y^3+7569y^2-5436y+797}{108(y-1)^4}\right] 
 \ - \f{4}{9} \, E_H                                   \nn \\[1.5ex]
 W_{8,\smallyy}   & = &
\frac{1}{6}\,y\left[
 \frac{13y^3-17y^2+30y}{(y-1)^4}\,
 {\rm Li}_2  \left(\!1\! -\! \f{1}{y} \right)
-\frac{17y^2+31y}{(y-1)^5}\ln^2y \right.               \nn \\
                  &  &
\phantom{ \frac{1}{6}\,y\left[\right.}
+\frac{42y^4+318y^3+1353y^2+817y-226}{36(y-1)^5}\ln y  \nn \\
                  &  &
\phantom{ \frac{1}{6}\,y\left[\right.}
\left.
+\frac{-4451y^3+7650y^2-18153y+1130}{216(y-1)^4}\right]
 \ - \f{1}{6} \, E_H                                   \nn \\[1.7ex]
 M_{7,\smallyy}   & = &
\frac{1}{27}y \, \left[
 \frac{-14y^4+149y^3-153y^2-13y+31-(18y^3+138y^2-84y) \ln y}{(y-1)^5}
 \right]                                               \nn \\[1.5ex]
 M_{8,\smallyy}   & = &
\frac{1}{36}y\left[
 \frac{-7y^4+25y^3-279y^2+223y+38+(102y^2+186y) \ln y}{(y-1)^5}
            \right]                                   \nn \\[1.7ex]
T_{7,\smallyy}    &=& \frac{1}{3} \, T_{7,\smallsm}(x \to y)
                                                      \nn \\[1.5ex]
T_{8,\smallyy}    &=& \frac{1}{3} \, T_{8,\smallsm}(x \to y) \,.
\label{wcnloyy}
\eea

\newpage 
\noindent
{\bf 2HDM case, coupling $(XY^\star)$}

\noindent
Similarly, the functions relative to the charged Higgs
contributions proportional to $({\mbox{\small X Y}}^\ast)$ are
at the LO:
\bea
 C_{7,\smallxy}^0 & = & \frac{1}{12} \, y \, \left[
 \frac{-5y^2+8y-3+(6y-4)\ln y}{(y-1)^3} \right]
                        \nn \\[1.5ex]
 C_{8,\smallxy}^0 & = & \frac{1}{4} \, y \, \left[
\frac{-y^2+4y-3- 2 \ln y}{(y-1)^3} \right] \,,
\label{wcloxy}                                             \\[1.7ex]
\fudge{and at the NLO:}                                    \\[1.1ex]
 W_{7,\smallxy}   & = &
\frac{4}{3}\,y\left[
 \frac{8y^2-28y+12}{3(y-1)^3} \,
 {\rm Li}_2 \left(\!1\! -\! \f{1}{y} \right)+
\frac{3y^2+14y-8}{3(y-1)^4}\ln^2y \right.            \nn \\
         &   &
\phantom{ \frac{4}{3}\,y\left[\right.}
\left. +\frac{4y^3-24y^2+2y+6}{3(y-1)^4}\ln y
+\frac{-2y^2+13y-7}{(y-1)^3}\right]
                                                      \nn \\[1.5ex]
 W_{8,\smallxy}   & = &
\frac{1}{3}\,y\left[
 \frac{17y^2-25y+36}{2(y-1)^3}\,
 {\rm Li}_2 \left(\!1\! -\! \f{1}{y} \right) -
  \frac{17y+19}{(y-1)^4}\ln^2y \right.              \nn \\
         &  &
\phantom{ \frac{1}{3}\,y\left[\right.}
\left. +\frac{14y^3-12y^2+187y+3}{4(y-1)^4}\ln y
-\frac{3(29y^2-44y+143)}{8(y-1)^3}\right]
                                                      \nn \\[1.7ex]
 M_{7,\smallxy}   & = &
 \frac{2}{9} \, y \, \left[
 \frac{-8y^3+55y^2-68y+21-(6y^2+28y-16) \ln y}{(y-1)^4}
 \right]
                                                      \nn \\[1.5ex]
 M_{8,\smallxy}   & = &
  \frac{1}{6} \, y \, \left[
 \frac{-7y^3+23y^2-97y+81 +(34y+38) \ln y}{(y-1)^4}
                    \right]                           \nn \\[1.7ex]
T_{7,\smallxy}    &=&
 \frac{2}{3} \, y \, \left[ \frac{13y^2-20y+7-(6y^2+4y-4)
                       \ln y}{(y-1)^4} \right]       \nn \\[1.5ex]
T_{8,\smallxy}    &=& 2y \, \left[ \frac{-y^2-4y+5+(4y+2)
                       \ln y}{(y-1)^4} \right] \,.
\label{wcnloxy}
\eea

\section{Anomalous Dimension Matrix}
\label{anomalousdm}
\noindent 
For completeness, we give the 
anomalous dimension matrix which govern 
the evolution of the Wilson Coefficients 
from $\muw$ to $\mub$. 
%$\gamma_{ji}^{\,{\rm eff}}(\mu)$ in~(\ref{RGE})
It can be expanded perturbatively as:
\beq 
 \gamma_{ji}^{\,{\rm eff}}(\mu)   = 
   \f{\as (\mu)}{4 \pi}  \,
 \gamma_{ji}^{0,\,{\rm eff}}
 + \f{\as^2(\mu)}{(4 \pi)^2} \, 
 \gamma_{ji}^{1,\,{\rm eff}} + {\cal O}(\as^3)\,.
\eeq
The matrix  $\gamma_{ji}^{0,\,{\rm eff}}$ is given by:
\beq 
\label{gamma0}
 \left\{\gamma_{ji}^{0,\,{\rm eff}}\right\}  = \left[
\begin{array}{rrrrrrrr}
\vspace{0.2cm}
-4       & \f{8}{3}  &       0       &   -\f{2}{9}    &      
 0       &     0     & -\f{208}{243} &  \f{173}{162} \\ 
\vspace{0.2cm}
12       &     0     &       0       &    \f{4}{3}    &     
 0       &     0     &   \f{416}{81} &    \f{70}{27} \\ 
\vspace{0.2cm}
 0       &     0     &       0       &  -\f{52}{3}    &    
 0       &     2     &  -\f{176}{81} &    \f{14}{27} \\ 
\vspace{0.2cm}
 0       &     0     &  -\f{40}{9}   & -\f{100}{9}    & 
\f{4}{9} &  \f{5}{6} & -\f{152}{243} & -\f{587}{162} \\ 
\vspace{0.2cm}
 0       &     0     &       0       & -\f{256}{3}    &     
 0       &    20     & -\f{6272}{81} &  \f{6596}{27} \\ 
\vspace{0.2cm}
 0       &     0     & -\f{256}{9}   &   \f{56}{9}    & 
\f{40}{9}& -\f{2}{3} & \f{4624}{243} &  \f{4772}{81} \\ 
\vspace{0.2cm}
 0       &     0     &       0       &       0        &     
 0       &     0     &     \f{32}{3} &       0       \\ 
\vspace{0.2cm}
 0       &     0     &       0       &       0        &     
 0       &     0     &    -\f{32}{9} &  \f{28}{3}    \\
\end{array} \right] \,, 
\eeq
and in the $\overline{MS}$ scheme with fully anticommuting 
$\gamma_5$, 
$ \gamma_{ji}^{1,\,{\rm eff}}$ is~\cite{CMM}:
\beq 
\label{gamma1}
 \left\{\gamma_{ji}^{1,\,{\rm eff}}\right\}  = \left[
\begin{array}{rrrrrrrr}
\vspace{0.2cm}
-\f{355}{9}    & -\f{502}{27}  & -\f{1412}{243}   & -\f{1369}{243}   &    
 \f{134}{243}  &  -\f{35}{162} & -\f{818}{243}    &  \f{3779}{324}   \\ 
\vspace{0.2cm}
 -\f{35}{3}    &   -\f{28}{3}  & -\f{416}{81}     &  \f{1280}{81}    &     
 \f{56}{81}    &   \f{35}{27}  &  \f{508}{81}     &  \f{1841}{108}   \\ 
\vspace{0.2cm}
     0         &        0      & -\f{4468}{81}    & -\f{31469}{81}   &    
 \f{400}{81}   & \f{3373}{108} &  \f{22348}{243}  &  \f{10178}{81}   \\ 
\vspace{0.2cm}
     0         &        0      & -\f{8158}{243}   & -\f{59399}{243}  &   
 \f{269}{486}  & \f{12899}{648}& -\f{17584}{243}  & -\f{172471}{648} \\ 
\vspace{0.2cm}
     0         &        0      & -\f{251680}{81}  & -\f{128648}{81}  &  
 \f{23836}{81} &  \f{6106}{27} & \f{1183696}{729} & \f{2901296}{243} \\ 
\vspace{0.2cm}
     0         &        0      &  \f{58640}{243}  & -\f{26348}{243}  & 
-\f{14324}{243}& -\f{2551}{162}& \f{2480344}{2187}& -\f{3296257}{729}\\ 
\vspace{0.2cm}
     0         &        0      &         0        &         0        &
     0         &        0      & \f{4688}{27}     &         0        \\  
\vspace{0.2cm}
     0         &        0      &         0        &         0        &
     0         &        0      & -\f{2192}{81}    &  \f{4063}{27}    \\
\end{array} \right] \,. 
\eeq

\section{``Running'' numbers}
\label{runnnumbers}
\noindent 
The vectors $\{a_i\}$, $\{h_i\}$, and 
$\{a^\prime_i\}$,  $\{h^\prime_i\}$,
needed for the evaluation of the low--scale Wilson coefficients
$C_7^{0,\,{\rm eff}}(\mub)$ and 
$C_8^{0,\,{\rm eff}}(\mub)$ are:
\bea
 \{a_i\} & = &
 \left\{ 
   \, \f{14}{23}, \,\f{16}{23},\, \f{6}{23}, -\f{12}{23}, \,0.4086, 
    -0.4230, -0.8994, \,0.1456 
 \right\}                                       \nn \\
 \{h_i\} & = &
 \left\{ 
   \, \f{626126}{272277}, -\f{56281}{51730}, -\f{3}{7}, -\f{1}{14},
   -0.6494, -0.0380, -0.0186, -0.0057             
 \right\}\,,
\label{lo7runn}                                         \\[1.1ex]
 \{a^\prime_i\} & = &
 \left\{ 
   \, \f{14}{23}, \,0.4086, 
    -0.4230, -0.8994, \,0.1456 
 \right\}                                       \nn \\
 \{h^\prime_i\} & = &
 \left\{ 
   \, \f{313063}{363036}, 
   -0.9135, 0.0873, -0.0571, 0.0209             
 \right\}\,;
\label{lo8runn}                                         \\[1.1ex]
\fudge{those
needed for $C_7^{1,\,{\rm eff}}(\mub)$:}       \\    
 \{e_i\} & = &
 \left\{ 
  \, \f{4661194}{816831}, -\f{8516}{2217}, \,0, \,0, -1.9043,
   -0.1008, \,0.1216, \,0.0183
 \right\}  \quad                                \nn \\
 \{f_i\} & = &
 \left\{ 
   -17.3023, \,8.5027, \,4.5508, \,0.7519, \,2.0040, \,0.7476,        
   -0.5385, \,0.0914         
 \right\}  \quad                                \nn \\
 \{k_i\} & = &
 \left\{ 
   \, 9.9372,-7.4878, \,1.2688, -0.2925,-2.2923,
   -0.1461, \,0.1239, \,0.0812
 \right\}  \quad                                \nn \\
 \{l_i\} & = &
 \left\{ 
   \, 0.5784, -0.3921, -0.1429, \, 0.0476,
   -0.1275, \,0.0317, \,0.0078, -0.0031
 \right\} \,. \quad
\label{nlorunn} 
\eea

\section{Virtual correction functions $r_i$}
\label{specifyD}
\noindent 
The renormalization scale independent parts of the 
virtual corrections, encoded in the functions $r_i$
in eq.~(\ref{vdefine}), read:
\bea
 r_1 & = & -\f{1}{6} r_2                                  \nn \\ 
 r_2 & = &
  \f{2}{243} \,\left\{-833+144 \pi^2 z^{3/2}
               \right.                                    \nn \\
     &   &  \hspace{1.2cm}
+ \left[ 1728 -180 \pi^2 -1296 \,\zeta (3)
       +(1296 -324 \pi^2) L +108 L^2 +36 L^3
  \right]  z                                              \nn \\
     &   &  \hspace{1.2cm}
+ \left[ 648 +72 \pi^2 +(432 -216 \pi^2) L +36 L^3
  \right]    z^2                                          \nn \\
     &   &  \hspace{1.2cm} 
 \left.                 +
 \left[ -54 -84 \pi^2 +1092 L -756 L^2
 \right]    z^3  \,
 \right\}                                                 \nn \\
     &   &
 + \f{16 \pi i}{81} \, \left\{ -5 
 + \left[ 45 -3 \pi^2 + 9 L + 9 L^2 \right]    z
 + \left[ -3 \pi^2 + 9 L^2 \right]    z^2
 + \left[ 28 - 12 L  \right]  z^3 \, \right\}             \nn \\
     &   &  \hspace*{1.2cm}
\hspace{0.3cm} + \hspace{0.3cm} {\cal O}(z^4)             \nn \\
 r_7 & = &
   \f{32}{9} -\f{8}{9} \pi^2                              \nn \\
 r_8 & = &
 -\f{4}{27} ( -33 + 2 \pi^2 - 6 i \pi )\,,
\eea
where $z$ is defined as $z=m_c^2/m_b^2$ and the symbol
$L$ denotes $L=\ln(z)$. 
Notice that $r_3\,$,$r_4\,$,$r_5$, and $r_6$ are not used 
in the approximation $C_i^{0,\,{\rm eff}}(\mub)=0$ ($i=3,4,5,6$)
used for the matrix elements.

\section{Bremsstrahlung $f_{ij}$ terms}
\label{defineeffij}
\noindent 
The expressions for the bremsstrahlung functions $f_{ij}$ we use
in the present paper are
obtained after integrating one variable in the expressions 
given in Appendix B in ref. \cite{GHW}. The explicit expressions
read (converted to the operator basis \ref{opbasis}):
\begin{eqnarray}
\label{fs}
 f_{11} &=&\phantom{-} \frac{1}{36}\, f_{22}        \nn \\
 f_{12} &=& -\frac{1}{3}\, f_{22}                   \nn \\
 f_{17} &=& -\frac{1}{6}\, f_{27}                   \nn \\ 
 f_{18} &=& -\frac{1}{6}\, f_{28}                   \nn \\[1.5ex]
 f_{22} &=&\phantom{-}\frac{16z}{27} \,
  \int_{0}^{1/z} \, dt (1-zt)^2 \, \left\vert 
  \frac{G(t)}{t} + \frac{1}{2} \right\vert^2          \nn \\
 f_{27} &=& - \frac{8z^2}{9} \,
  \int_{0}^{1/z} \, dt (1-zt)\, \left( 
  G(t) + \frac{t}{2} \right)                      \nn \\
 f_{28} &=& - \frac{1}{3} \, f_{27}               \nn \\[1.5ex]
 f_{78} &=& \,\frac{8}{9} \, 
  \left( \frac{25}{12} - \frac{\pi^2}{6} \right)  \nn \\[1.5ex]
 f_{88} &=& \frac{1}{27} \,
  \left( \frac{16}{3} - \frac{4\pi^2}{3} + 4 \ln 
  \frac{m_b}{\mu_b} \right) \,,
\end{eqnarray}
where $z$ is $z=m_c^2/m_b^2$ and $\mu_b$ is the renormalization scale.
The function $G(t)$ appearing in eq.~(\ref{fs}) reads
\beq
 G(t) =
 \left\{ \begin{array}{cc} 
  - 2 \arctan^2 \sqrt{\f{t}{4-t}}                              &   
\mbox{ for $t < 4$} \vspace{0.2cm}                            \\[1.ex]
  -\pi^2/2 + 2 \ln^2\left(\f{\sqrt{t} + \sqrt{t-4}}{2}\right) 
  - 2 i \pi \ln\left(\f{\sqrt{t} + \sqrt{t-4}}{2}\right)       &
\mbox{ for $t \geq 4$}. 
\end{array} \right.
\eeq

\section{Input Parameters}
\label{InputsScales}
\noindent 
We list in table~\ref{inputpar}
the values of input parameters used in our calculation. 
\begin{table}[h]
\small{
\begin{center}
\begin{tabular}{|c|c|c|c|c|c|c|}
\hline
 $\as(\mz)$  & $m_{t}$ & $m_{c}/m_{b}$ & $m_{b}-m_{c}$ &
 $\alpha_{em}^{-1}$ &  $|V_{ts}^\star V_{tb}/V_{cb}|^2$ &
 ${\rm BR}_{\smallSL}$                       \\
             & (GeV) & & (GeV) & & &
                                 \\
\hline
 $\phantom{\pm} 0.119$ & $\phantom{\pm} 175$    &
 $\phantom{\pm} 0.29$  &
 $\phantom{\pm} 3.39$  & $\phantom{\pm} 130.3$  &
 $\phantom{\pm} 0.95$  & $\phantom{\pm} 0.1049$           \\
\hline
  $\pm 0.004 $ & $\pm  5.0 $ 
 &$\pm 0.02  $ & $\pm 0.04 $ 
 &$\pm  2.3  $ & $\pm 0.03 $  &$\pm 0.0046$     \\
\hline
\end{tabular} 
\caption{{\small Central value of our input parameters (first line)
and their uncertainties (second line).}}
\label{inputpar}
\end{center}
}
\end{table}
The masses $m_t$, $m_b$, and $m_c$ are understood to be the pole
masses of the top, bottom and charm quark. The 
value of the $m_t$ is obtained by combining the results 
given in~\cite{CDF,D0}; those of the two combinations
of $m_c$ and $m_b$ are taken from~\cite{BIGI}.
For $\alpha_{em}$ and 
$|V_{ts}^\star V_{tb}/V_{cb}|^2$ we refer to~\cite{ALPHAEM}
and~\cite{PDG}, respectively. 
We take 
$\as(\mz) = 0.119 \pm 0.004$, as an average between a pessimistic
and optimistic estimate of the error~\cite{ALPHASBE}, as 
suggested in~\cite{ALPHASAL}. The value of the semileptonic
branching ratio ${\rm BR}_{\smallSL}$ has been recently obtained 
by the CLEO Collaboration~\cite{CLEOSL}.         
The other 
constants used in the calculation are $\mw=80.33$ GeV, 
$\lambda_1=-0.5$ GeV$^2$, and $\lambda_2=0.12$ GeV$^2$.


\newpage
\small 
\begin{thebibliography}{99}

\bibitem{Bbook}
{\sc S. Bertolini, F. Borzumati, and A. Masiero},
{\it Probing New Physics in FCNC B Decays and Oscillations}, 
in {\it ``B Decays''}, S. Stone~(Ed.),  
World Scientific, Singapore, 1992, pp. 458--478 (1st Edition), 
and 1994, pp. 620--643  (2nd Edition)   

\bibitem{H-BBP}
{\sc  J.L. Hewett}, \prl{70}{93}{1045}; \\ 
{\sc  V. Barger, M. Berger, and R.J.N. Phillips}, \prl{70}{93}{1368}

\bibitem{SUSYHiggs}
{\sc S.~Dimopoulos and H.~Georgi}, \npb{193}{81}{150}

\bibitem{IO}
 For a discussion within the Minimal Supersymmetric 
 Standard Model see:                   \\
{\sc F. Borzumati}, \zpc{63}{94}{291}; \\
{\sc F. Borzumati and N. Polonsky},                    
 Contribution to {\it $e^+e^-$ Collisions at TeV Energies: The 
 Physics Potential}, Annecy, Gran Sasso, Hamburg,
 February--September 1995, P.M.~Zerwas (Ed.), hep--ph/9602433

\bibitem{CLEO}
{\sc  M.S.~Alam} {\it et al.} (CLEO Collab.), \prl{74}{95}{2885}

\bibitem{ALEPH}
as reported by {\sc P.M.~Kluit}, talk 906 at the International 
Europhysics Conference on High Energy Physics 1997

\bibitem{BBM}
{\sc  S. Bertolini, F. Borzumati, and A. Masiero},
  \prl{59}{87}{180}; \\
{\sc  N.G. Deshpande, P. Lo, J. Trampetic, G. Eilam, and P. Singer},
  \prl{59}{87}{183}

\bibitem{GSW}
{\sc  B. Grinstein, R. Springer, and M.B. Wise}, 
  \plb{202}{88}{138}; \npb{339}{90}{269}

\bibitem{CIUCHINI2}
{\sc  M.~Ciuchini, E.~Franco, G.~Martinelli, L.~Reina, 
  and L.~Silvestrini}, \plb{316}{93}{127};\\
{\sc  M.~Ciuchini, E.~Franco, L.~Reina, and L.~Silvestrini},
  \npb{421}{94}{41};\\
{\sc  M. Ciuchini, E. Franco, G. Martinelli, and L. Reina},
  \plb{301}{93}{263}, \npb{415}{94}{403};\\
{\sc M. Misiak},  \npb{393}{93}{23}; E.~\ib{439}{95}{461};  \\
{\sc G.~Cella, G.~Curci, G.~Ricciardi, and  A.~Vicer{\'e}},
 \plb{325}{94}{227}; \npb{431}{94}{417}

\bibitem{AG}
{\sc  A.~Ali and C.~Greub}, \zpc{49}{91}{431}; \plb{259}{91}{182};
 \plb{361}{95}{146}

\bibitem{LOWSCDEP}
{\sc A. Ali and C. Greub}, \zpc{60}{93}{433}

\bibitem{CIUCHINI1}
{\sc  M. Ciuchini, E. Franco, G. Martinelli, and L. Reina},
  \plb{334}{94}{137}

\bibitem{NORESULTS}
{\sc  A.J. Buras, M. Misiak, M. M\"unz, and S. Pokorski},
 \npb{424}{94}{374}

\bibitem{AYAO}
{\sc  K.~Adel and Y.P.~Yao}, \prd{49}{94}{4945}

\bibitem{cGtH}
{\sc  C.~Greub and T.~Hurth}, \prd{56}{97}{2934}

\bibitem{INFRARED}
{\sc A.J. Buras, A. Kwiatkowski, and N. Pott}, hep-ph/9710336  

\bibitem{CDGG}
{\sc M.~Ciuchini, G.~Degrassi, P.~Gambino, and G.F.~Giudice},
  hep-ph/9710335

\bibitem{GHW}
{\sc  C.~Greub, T.~Hurth, and D.~Wyler}, \plb{380}{96}{385}; 
  \prd{54}{96}{3350}

\bibitem{BURAS}
{\sc  A.J. Buras, M. Jamin, M.E. Lautenbacher, and P.H. Weisz},
  \npb{370}{92}{69}; \npb{375}{92}{501} (addendum)

\bibitem{MISIAK95}
{\sc M. Misiak and M. M{\"u}nz}, \plb{344}{95}{308}

\bibitem{CMM}
{\sc  K.~Chetyrkin, M.~Misiak, and M.~M{\"u}nz}, \plb{400}{97}{206}

\bibitem{CMMlong}
{\sc  K.~Chetyrkin, M.~Misiak, and M.~M{\"u}nz}, hep-ph/9711266 

\bibitem{FALK}
{\sc  A. Falk, M. Luke, and M. Savage}, \prd{49}{94}{3367}

\bibitem{BIGI}
{\sc  I.I. Bigi, M. Shifman, N.G. Uraltsev, and A.I. Vainshtein},
 \prl{71}{93}{496}

\bibitem{LUKE}
{\sc  A.V. Manohar and M.B. Wise}, \prd{49}{94}{1310};        \\
{\sc  A. Falk, M. Luke, and M. Savage}, \prd{53}{96}{2491}

\bibitem{VOLOSHIN}
{\sc  M.B. Voloshin}, \plb{397}{97}{295};                    \\
{\sc  Z. Ligeti, L. Randall, and M.B. Wise}, \plb{402}{97}{178};\\
{\sc  A.K. Grant, A.G. Morgan, S. Nussinov, and R.D. Peccei},
  \prd{56}{97}{3151};                                           \\ 
{\sc  G. Buchalla, G. Isidori, and S.J. Rey}, hep-ph/9705253    

\bibitem{BKP}
{\sc A.J. Buras, A. Kwiatkowski, and N. Pott}, \plb{414}{97}{157}

\bibitem{HW}
{\sc  W.S. Hou and R.S. Willey}, \plb{202}{88}{591}

\bibitem{BHP}
{\sc  V. Barger, J.L. Hewett, and R.J.N. Phillips}, \prd{41}{90}{3421}

\bibitem{BBMR}
{\sc  S. Bertolini, F. Borzumati, A. Masiero, and G. Ridolfi}, 
  \npb{353}{91}{591}

\bibitem{KP}
{\sc P. Krawczyk and S. Pokorski}, \npb{364}{91}{11}

\bibitem{GNR}
{\sc Y. Grossman and Y. Nir}, \plb{313}{93}{126}; \\
{\sc Y. Grossman, Y. Nir, and R. Rattazzi}, to appear 
 in ``Heavy Flavour II'', A.J. Buras and M. Lindner (Eds.),
 hep-ph/9701231

\bibitem{CRS}
{\sc P.~Ciafaloni, A.~Romanino, and A.~Strumia}, hep-ph/9710312

\bibitem{GW} 
{\sc S.L. Glashow and S. Weinberg}, \prd{15}{77}{1958}

\bibitem{WEIN}
{\sc S. Weinberg}, \prl{37}{76}{657}

\bibitem{AST}
{\sc C.H. Albright, J. Smith, and S.-H.H. Tye}, \prd{21}{80}{711}

\bibitem{BBG}
{\sc G.C. Branco, A.J. Buras, and J.M. Gerard}, \npb{259}{85}{306}

\bibitem{WEI}
{\sc S. Weinberg}, \prd{42}{90}{860}

\bibitem{LAV}
{\sc L. Lavoura}, \ijmpa{8}{93}{375}

\bibitem{HHS}
{\sc  H.E.~Haber, G.L.~Kane, and T.~Sterling}, \npb{161}{79}{493} 

\bibitem{POLSIMMA}
{\sc  H. D. Politzer}, \npb{172}{80}{349}; \\
{\sc  H.~Simma}, \zpc{61}{94}{67}

\bibitem{INAMILIM}
{\sc  T. Inami and C.S. Lim}, \ptp{65}{81}{297}

\bibitem{BBL}
{\sc G. Buchalla, A.J. Buras, and M. Lautenbacher}, \rmp{68}{96}{1125}

\bibitem{POTT}
{\sc  N.~Pott}, \prd{54}{96}{938}

\bibitem{CABIB}
{\sc  N. Cabibbo and L. Maiani}, \plb{79}{78}{109};\\
{\sc  C.S. Kim and A.D. Martin}, \plb{225}{89}{186};\\
{\sc  Y. Nir}, \plb{221}{89}{184}

\bibitem{GHHawaii}
{\sc  C.~Greub and T.~Hurth}, hep-ph/9708214, to appear in:
Proceedings of the Second Int. Conf. on B Physics and CP Violation,
Honolulu, Hawaii, 24-27 March 1997. 

\bibitem{BD}
{\sc F.M.~Borzumati and A.~Djouadi}, in preparation

\bibitem{AAG}
{\sc A.~Ali, Asatrian, and C.~Greub}, in preparation

\bibitem{MMM}
{\sc F.M.~Borzumati}, hep-ph/9702307  

\bibitem{CDF}
{\sc  S. Leone}, for the CDF Coll., Proc. 
 High--Energy Physics International Euroconference
 on Quantum Chromodynamics: QCD 97, Montpellier, France, 3-9 Jul
 1997

\bibitem{D0}
{\sc  K. Genser}, for the D0 Coll., Proc.
 11th Les Rencontres de Physique de la Vallee d'Aoste: Results and
 Perspectives in Particle Physics, La Thuile, Italy, 2-8 Mar 1997

\bibitem{ALPHAEM}
{\sc S.~Eidelman and F.~Jegerlehner}, \zpc{67}{95}{585}

\bibitem{PDG}
{\sc R.M. Barnett} {\it et al.} (PARTICLE DATA GROUP),
 \prd{54}{96}{1}

\bibitem{ALPHASBE}
{\sc S. Bethke,} {\it QCD Tests at $e+e-$ Colliders}
 talk given at High--Energy Physics International Euroconference 
 on Quantum Chromodynamics: QCD 97: 25th Anniversary of QCD,
 Montpellier, France, 3-9 Jul 1997, hep-ph/9710030 
 
\bibitem{ALPHASAL}
{\sc G. Altarelli},  {\it The Status of the Standard Model},
 talk given at the XVIII International Symposium on Lepton--Photon 
 Interactions, Hamburg, 1997, hep-ph/9710434 

\bibitem{CLEOSL}
{\sc  B. Barish} {\it et al.} (CLEO Coll.), \prl{76}{96}{1570}

\end{thebibliography}
\end{document}